\documentclass[12pt]{elsarticle}
\usepackage{amsmath}
\usepackage{amssymb}
\usepackage{mathtools}
\usepackage{color}
\usepackage{graphicx}
\usepackage{theorem}
\usepackage{topcapt}
\usepackage{caption}
\usepackage{subcaption}
\usepackage{float}
\usepackage{array}
\usepackage{tikz}
\usepackage{pgfplots}
\usepackage{pgfplotstable}
\usepackage{filecontents}
\usepackage{multirow}
\usepackage{afterpage}
\usepackage[hyphens]{url}
\usepackage{natbib}
\usetikzlibrary{shapes,arrows}
\tikzstyle{decision} = [diamond, draw, fill=blue!20,
    text width=4.5em, text badly centered, node distance=3cm, inner sep=0pt]
\tikzstyle{block} = [rectangle, draw, fill=blue!20,
    minimum width=6em, text centered, rounded corners, minimum height=4em]
\tikzstyle{line} = [draw, -latex']
\tikzstyle{cloud} = [draw, ellipse,fill=red!20, node distance=3cm,
    minimum height=2em]

\numberwithin{equation}{section}

\biboptions{comma, square, sort&compress}

\journal{Journal of Computational Physics}

\newcommand{\norm}[1]{\|#1\|}

\begin{document}

\begin{frontmatter}

\title{A Parallel Multi-Domain Solution Methodology Applied to Nonlinear Thermal Transport Problems in Nuclear Fuel Pins\tnoteref{tnote}}

\tnotetext[tnote]{Notice: This manuscript has been authored by UT-Battelle, LLC, under Contract No. DE-AC05-00OR22725 with
 the U.S. Department of Energy. The United States Government retains and the publisher, by accepting the
 article for publication, acknowledges that the United States Government retains a non-exclusive,
 paid-up, irrevocable, world-wide license to publish or reproduce the published form of this
 manuscript, or allow others to do so, for United States Government purposes.
}

\author[ornl]{Bobby Philip\corref{cor1}}
\author[ornl]{Mark A. Berrill}
\author[ornl]{Srikanth Allu}
\author[ornl]{Steven P. Hamilton}
\author[ornl]{Rahul S. Sampath}
\author[ornl]{Kevin T. Clarno}
\author[lanl]{Gary A. Dilts}

\cortext[cor1]{Corresponding author: Bobby Philip (Email: philipb@ornl.gov)}

\address[ornl]{Oak Ridge National Laboratory \\ One Bethel Valley Road, Oak Ridge, TN 37831}
\address[lanl]{Los Alamos National Laboratory \\ P.O. Box 1663, Los Alamos NM 87545}

\begin{abstract}

This paper describes an efficient and nonlinearly consistent parallel solution methodology for solving coupled nonlinear thermal transport problems that occur in nuclear reactor applications over hundreds of individual 3D physical subdomains. Efficiency is obtained by leveraging knowledge of the physical domains, the physics on individual domains, and the couplings between them for preconditioning within a Jacobian Free Newton Krylov method. Details of the computational infrastructure that enabled this work, namely the open source Advanced Multi-Physics (AMP) package developed by the authors is described. Details of verification and validation experiments, and parallel performance analysis in weak and strong scaling studies demonstrating the achieved efficiency of the algorithm are presented. Furthermore, numerical experiments demonstrate that the preconditioner developed is independent of the number of fuel subdomains in a fuel rod, which is particularly important when simulating different types of fuel rods. Finally, we demonstrate the power of the coupling methodology by considering problems with couplings between surface and volume physics and coupling of nonlinear thermal transport in fuel rods to an external radiation transport code.
\end{abstract}


\end{frontmatter}

\section{Introduction}
\label{sec:intro}

 Many real world engineering problems involve multiple coupled nonlinear physical processes that occur both within and across several
 interacting physical domains. Robust, accurate, and efficient three dimensional simulations for some of these complex problems pose significant challenges that require a
 combination of powerful numerical algorithms, efficient parallel implementations, and massive computing resources to tackle. These challenges include developing the numerical methods and the parallel software infrastructure for coupling physical phenomena that occur on the surface and within the interior of physical domains, 
 coupling structured and unstructured mesh calculations, coupling models with different discretizations, and using tightly coupled solution methods to 
 solve certain coupled physics problems and loosely coupled approaches for others. Developing such simulation capabilities
 is a nontrivial task.
 
 In this article, we will focus primarily on one such complex application where all of the features outlined above are present: thermal transport in nuclear fuel rods. However, we will also devote some effort to describing the parallel code infrastructure that was developed to provide the necessary meshing, discretization, linear algebra, linear and nonlinear solvers, physics modules (conservation laws and constitutive models), material property databases, and parallelization mechanisms for simulating this application in hopes that it will be beneficial to the broader scientific community.
 
\begin{figure}
   \centering
   \includegraphics[scale=0.2]{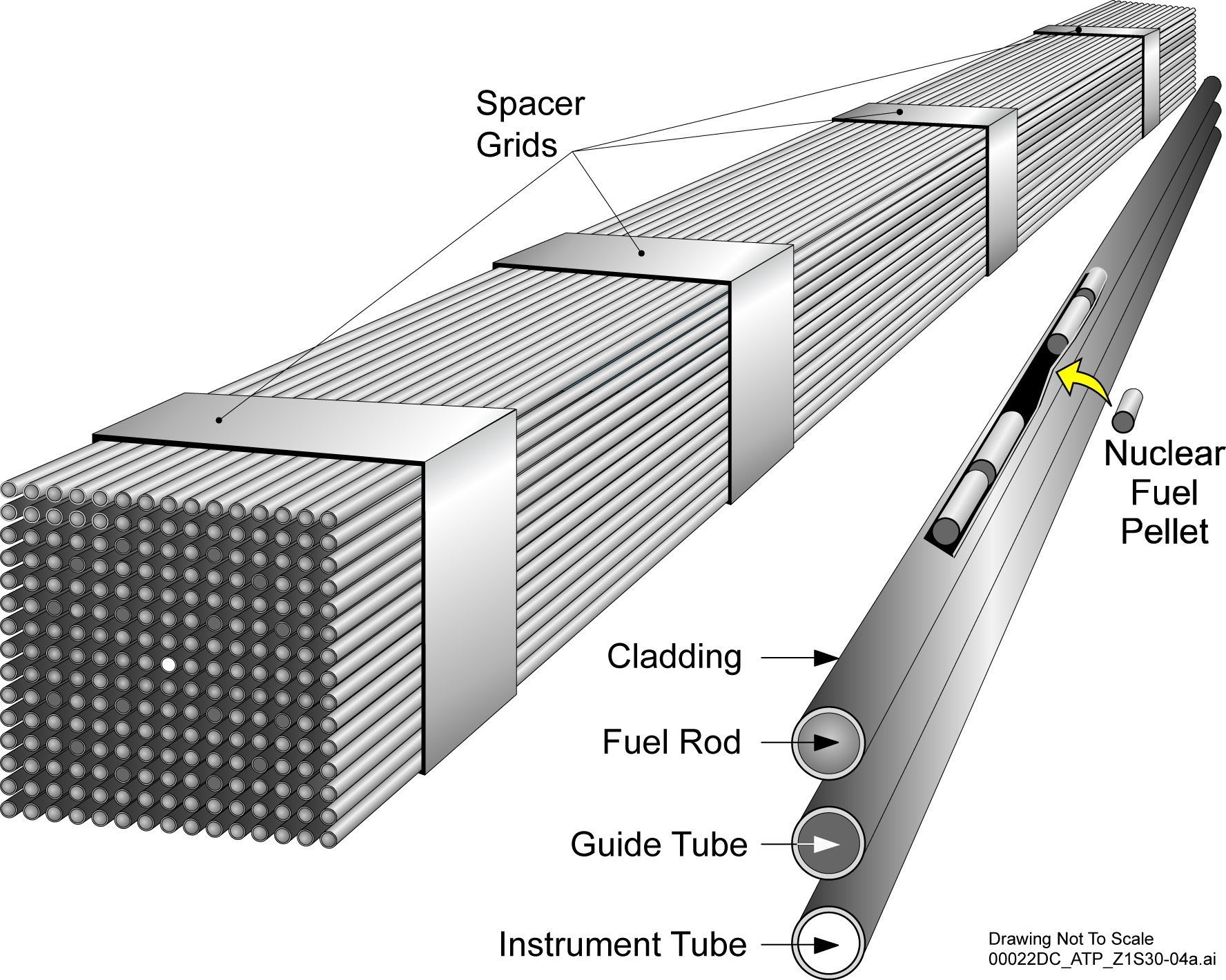}
   \caption{ Fuel assembly containing nuclear fuel rods that are filled with fuel pellets \cite{doe_yucca_website}  } 
   \label{fig:yuccaFig}
 \end{figure}

 A nuclear fuel assembly consists of several hundred nuclear fuel rods (shown in Figure \ref{fig:yuccaFig}) bound together by spacer grids.
While some of the rod locations are reserved for instrumentation and safety, most of the rods contain nuclear fuel.  Each individual
 nuclear fuel rod in turn consists of several hundred nearly cylindrical nuclear fuel pellets (each with a height to diameter ratio of approximately
 one) stacked one on top of another to form a long column enclosed within a metal tube called the clad. Heat is generated
 within the pellets by nuclear fission and is distributed within the pellets and clad via a diffusive process. There is thermal 
 contact (modeled as a convective process) between neighboring pellets and between the pellets and the clad. Each fuel rod is 
 cooled with water flowing axially up the outer surface of the clad.  
 
 Modeling the heat transfer, along with other physics, leads to a very high-aspect ratio problem with many inter-dependent domains.
 Traditional nuclear fuel simulation eliminates the computational challenge by approximating the heat transfer as entirely radial 
 and neglecting the axial and azimuthal components, which are only coupled through the coolant temperature and other simplified 
 physics \cite{frapcon_manual, falcon_manual}. Recent efforts to develop advanced modeling and simulation tools for nuclear fuel
 rods \cite{newman_09, thouyenin_07}, which include simulating full three-dimensional fuel rods with resolved pellets, have relied 
 upon standard solution and preconditioning strategies that do not necessarily take 
 advantage of the physics and geometry of the problem.
 However, this manuscript does not address the challenges of structural dynamics and the associated feedback on heat transfer.

 With respect to specific work related to heat transfer within nuclear fuel rods, there are several existing efforts to develop 
 parallel codes that model three-dimensional heat transfer within nuclear fuel rods, including PLEIADES/ALCYONE \cite{thouyenin_06,
 newman_09}, MOOSE/Bison \cite{newman_09}, and BACO \cite{marino_07}. These codes are all focused on the integration of the many 
 physics required for modeling nuclear fuel performance in steady-state and transients to improve the underlying material science,
 including fracture/contact mechanics, fission gas generation and release, and corrosion chemistry. 

 We have developed an efficient scalable parallel simulation framework for solving such multi-domain, multi-physics problems
 and have used it to solve the specific nuclear fuel problem described above. Within our particular application a nonlinearly consistent Jacobian Free Newton Krylov (JFNK) method is used (though the ability to use alternative solution methods also exists) across all the domains for each fuel rod. Physics-based preconditioning is used to accelerate
 the solution process and the multi-domain (pellets and clad) aspect of the problem is leveraged in developing methods that minimize communication
 as well as avoid the formation of full matrices over the whole domain. 
 
 In the next section, we present a mathematical description of the problem under consideration. Section \ref{sec:discretization} will describe the finite element discretization of the models in Section \ref{sec:model}. The algorithms
 used to solve the resulting nonlinear system of equations are described in Section \ref{sec:algo}. The computational framework that was used in 
 this work is briefly described in Section \ref{sec:infrastructure}. Section \ref{sec:results} reports on numerical experiments performed to 
 verify and validate our code and test its parallel scalability. Section \ref{sec:assembly} provides details on coupling to reduced order flow models, coupling to oxide growth models on the exterior clad, and parallel full assembly simulations that couple thermal transport components on unstructured meshes with a structured mesh radiation transport code. The paper ends with a few concluding remarks.

\section{Model}
\label{sec:model}
A 3D fuel rod domain, $\Omega$, is modeled as consisting of the union of $N$ pellet subdomains, $\Omega^P_i \subset \mathbb{R}^3,\;i=1,2,\ldots, N$, and a clad subdomain $\Omega^C \subset \mathbb{R}^3$, i.e, the global domain $\Omega= \cup_{i=1}^N\Omega^P_i\cup \Omega^C$.
The number and geometric complexity of fuel pellets in fuel rods can vary significantly; from simple cylinders to the complex pellet geometries shown later in this manuscript and from a few pellets in an experimental rod to more than 400 pellets in a commerical nuclear fuel rod.
In our numerical experiments $N$ will be varied between $1$ and $360$  though there is no fundamental limitation on the number of pellet domains. Here, $\Omega_1^P$ will denote the domain of the lowest pellet and $\Omega_N^P$ will denote the topmost pellet domain. The surfaces of the pellet and clad domains will be denoted by $\Gamma^P_i, \; i=1, 2, \ldots, N$, and $\Gamma^C$, respectively. From now on, where there is no danger of confusion the superscript `P' will be dropped for the pellet subdomains. We will assume that the subdomains are disjoint. Only adjacent pellet subdomains are assumed to touch, i.e., $\Gamma_{i}\cap\Gamma_{i-1} \equiv \Gamma_{i,i-1} \neq \phi$ for $i=2, \ldots N$,  and $\Gamma_i\cap\Gamma_j=\phi$ for $j\neq i+1, i-1$. In this paper we will only consider the case where no pellets are in mechanical contact with the clad or flow subdomains. We will first describe the models at the level of individual pellets, clad, and coolant domains before considering the full coupled multi-domain model.

\subsection{Pellet Models: } The temperature field, $T_i$, in each fuel pellet domain $\Omega_i$ is modeled by a nonlinear thermal diffusion equation,
\begin{equation}
-\nabla\cdot k_i(T_i, \mathbf{x})\nabla{T_i}(\mathbf{x}) = f_i(\mathbf{x}), \; \text{for} \; \mathbf{x}\in \Omega_i,
\label{PelletThermalTransport}
\end{equation}
where $k_i$ is a scalar nonlinear thermal conductivity and $f_i \in L_2(\Omega_i)$ is a non-zero thermal source due to heat generated from nuclear fission in each pellet. In stand-alone applications, the radial shape of $f_i$ is usually approximated with a local exponential in the radial dimension and a globally low-order (quadratic or cosine) model in the axial dimension. $k_i$ and $f_i$ vary spatially within each pellet and can potentially vary in functional form from pellet to pellet to account for differences in materials and fission processes that do occur in practice. For example, insulator pellets with natural, rather than enriched, uranium are often introduced at the top and bottom of the pellet stack to reduce axial power peaking. 

There is contact/gap resistance between two adjacent pellets and for pellet $i$ this is modeled by Robin boundary conditions of the form 
\begin{eqnarray}
k_i(T_i)\nabla T_{i}\cdot \mathbf{n}_i + h_{i,i-1}(T_i, T_{i-1}^m) (T_{i} - T^m_{i-1} )  &=& 0 \; \text{on} \;\Gamma_{i,i-1}, \label{P2PRobinBC1}\\
k_i(T_i)\nabla T_{i}\cdot \mathbf{n}_i + h_{i,i+1}(T_i, T_{i+1}^m) (T_{i} - T^m_{i+1}) &=& 0 \; \text{on} \;\Gamma_{i,i+1}.
\label{P2PRobinBC2}
\end{eqnarray}
Here $ \mathbf{n}_i$ is the outward facing unit normal on the surface of pellet $i$ and $T^m_{i-1}$ is the surface temperature field for pellet $(i-1)$ on $\Gamma_{i-1}\cap\Gamma_{i,i-1}$  interpolated to the surface $\Gamma_{i}\cap\Gamma_{i,i-1}$ of pellet $i$. The flux between adjacent pellets is assumed to be continuous while the temperature field is assumed to be discontinuous.

Heat transfer between the pellets and clad is by radiative, conductive and convective processes through the gap region. For each pellet this is modeled using a Robin boundary condition
\begin{equation}
k_i(T_i)\nabla T_{i}\cdot \mathbf{n}_i + h_{i,c}(T_i, T^m_c) (T_{i} - T^m_{c} )  = 0 \; \text{on} \;\Gamma_{i,c}
\label{P2CRobinBC}
\end{equation}
where $T^m_c$ is the surface temperature field for the cladding projected onto the surface, $\Gamma_{i,c}$,  of pellet  $i$ and $h_{i,c}$ is an effective heat transfer coefficient that can be modeled to account for a wide variety of physical phenomena.  In this manuscript, we limit discussion of the heat transfer coefficient to the non-linearities associated with the dependence on the clad and fuel temperatures in the geometric orientation. This in effect results in a nonlinear coupling between all pellets and the clad through the boundary conditions. Zero Neumann boundary conditions 
\begin{equation}
k_i(T_i)\nabla T_{i}\cdot \mathbf{n}_i = 0 \; \text{on} \;\Gamma_{i,n}
\label{PNeumannBC}
\end{equation}
are imposed on all remaining surface boundary regions of each pellet.

\subsection{ Clad Model:} The temperature field, $T_c$, for the clad domain $\Omega_c$ 
is also modeled by a nonlinear thermal diffusion equation,
\begin{equation}
-\nabla\cdot k_c(T_c, \mathbf{x})\nabla{T_c}(\mathbf{x}) = 0 \;\text{for}\; \mathbf{x} \in \Omega_c
\label{CladThermalTransport}
\end{equation}
with a scalar nonlinear thermal conductivity ($k_c$) and a zero right hand side since 
heat is not generated within the clad materials. Heat transfer from the clad outer 
surface to the coolant is modeled by a Robin boundary condition
\begin{equation}
k_c(T_c)\nabla T_{c}\cdot \mathbf{n_f} + h_{c,f}(T_c, T^m_f) (T_{c} - T^m_{f} )  = 0 \; \text{on} \;\Gamma_{c,f} \\
\label{C2FRobinBC}
\end{equation}
where $\mathbf{n_f}$ is an outward facing unit normal from the clad surface into the flow region,
 $T^m_f$ denotes the interpolated temperature from the fluid flow region, and $h_{c,f}$ is the effective heat transfer coefficient. 
The temperature of the flow can be described either by a constant fixed temperature or
through an independent or coupled flow model.  For most of this work we choose to couple
a flow model that solves a form of the fluid equations as described in section \ref{sec:assembly} and \ref{sec:Coolant}.
Similarly, heat transfer between the clad inner surface and the gap region between 
the pellets and clad is modeled by a Robin boundary condition
\begin{equation}
k_c(T_c)\nabla T_{c}\cdot \mathbf{n_g} + h_{c,g}(T_c, T^m_1, \ldots, T^m_N) (T_{c} - T_{g} )  = 0 \; \text{on} \;\Gamma_{c,g} \\
\label{C2PRobinBC}
\end{equation}
Here $\mathbf{n_g}$ is the outward facing unit normal from the inner clad surface into the pellet-clad gap, $T^m_i$ is the temperature on the outer surface of the pellet projected onto the inner surface of the cladding for each pellet $i$, and $h_{c,g}$ is the clad-side heat transfer coefficient that is corresponds to the $h_{i,c}$ to conserve energy across the gap. In addition zero Neumann boundary conditions are imposed on the top and bottom of the clad cylinder.

\subsection{Weak Formulation:} 
\noindent Given $f_i(\mathbf{x}) \in L_2(\Omega_i)$, for each pellet subproblem we seek a solution to Eqns  (\ref{PelletThermalTransport})-(\ref{PNeumannBC}) in the trial space of functions
\begin{equation}
V_i = \{T(\mathbf{x})| \, \mathbf{x}\in \bar{\Omega}_i, \;T\in H^1(\Omega_i), \;T \;\text{satisfies Eqns (\ref{P2PRobinBC1})-(\ref{PNeumannBC}) on} \;\Gamma_i \}.
\end{equation}
The Galerkin weak formulation for each pellet subproblem is:
Find $T_i \in V_i$ such that $\forall v \in V_i$
\begin{equation}
- \int\limits_{\Omega_i} v\nabla\cdot k_i(T_i)\nabla{T_i} \, d\Omega_i = \int\limits_{\Omega_i} f_i v \, d\Omega_i.
\end{equation}
Integration by parts and the divergence theorem yield
\begin{equation}
\int\limits_{\Omega_i} k_i(T_i)\nabla{T_i}\cdot \nabla v \, d\Omega_i -  \int\limits_{\Gamma_i} v k_i(T_i)\nabla{T_i}\cdot \mathbf{n}_i \, d\Gamma_i= \int\limits_{\Omega_i} f_i v \, d\Omega_i 
\end{equation}
Using Eqns (\ref{P2PRobinBC1})-(\ref{PNeumannBC}) the boundary integral on the left hand side may be simplified to give
\begin{equation}
\begin{multlined}
 -\int \limits_{\Gamma_i} v k_i(T_i)\nabla{T_i}\cdot \mathbf{n}_i \, d\Gamma_i =\\
 \shoveleft[-0.4cm]{\int \limits_{\Gamma_{i,i-1}} \! \! v h_{i,i-1}(T_{i} - T^m_{i-1} )\, d\Gamma_i + \int\limits_{\Gamma_{i,i+1}} \! \! v h_{i,i+1}(T_{i} - T^m_{i+1} )\, d\Gamma_i +  \int\limits_{\Gamma_{i,c}} \! \! v h_{i,c}(T_{i} - T^m_{ic} )\, d\Gamma_i}
\end{multlined}
\end{equation}
Let  $\langle u,v\rangle _i = \int\limits_{\Omega_i} uv \,d\Omega_i$ and $(u,v)_{\gamma} = \int\limits_{\gamma} uv \,d\gamma $ denote the standard inner products over the domain $\Omega_i$ and a boundary segment $\gamma$ respectively. Then the Galerkin subproblems for the pellets, $i=1,2, \ldots, N$, can be written as:\\

\noindent {\bf Pellet Sub-Problems:} Find $T_i \in V_i$ such that $\forall v \in V_i$
\begin{equation}
\begin{multlined}
\shoveleft[-0.5cm]{\langle k_i(T_i)\nabla{T_i}, \nabla v\rangle _i + (h_{i,i-1}T_i, v)_{\Gamma_{i,i-1}} + (h_{i,i+1}T_i, v)_{\Gamma_{i,i+1}} + (h_{i,c}T_i, v)_{\Gamma_{i,c}}}\\
\shoveleft[-0.5cm]{= \;\langle f_i,v\rangle _i + (h_{i,i-1}T_{i-1}^m, v)_{\Gamma_{i,i-1}}+(h_{i,i+1}T_{i+1}^m, v)_{\Gamma_{i,i+1}}+(h_{i,c}T_c^m, v)_{\Gamma_{i,c}}}
\end{multlined}
\label{PelletWeakFormulation}
\end{equation}
Similarly, the Galerkin weak formulation for the clad subdomain may be stated as:\\

\noindent {\bf Clad Sub-Problem:} Find $T_c \in V_c$ such that $\forall v \in V_c$
\begin{equation}
\begin{multlined}
\shoveleft[-0.75cm]{\langle k_c(T_c)\nabla{T_c}, \nabla v\rangle _c + (h_{c,g}T_c, v)_{\Gamma_{c,g}} + (h_{c,f}T_c, v)_{\Gamma_{c,f}}}\\
\shoveleft[-0.7cm]{= \;\langle f_c,v\rangle _c + (h_{c,g}T_{g}^m, v)_{\Gamma_{c,g}}+(h_{c,f}T_{f}^m, v)_{\Gamma_{c,f}}}
\end{multlined}
\label{CladWeakFormulation}
\end{equation}
%
%

\section{Discretization}
\label{sec:discretization}
\subsection{Domain Discretization}

\begin{figure}[ht]%
    \centering
    \includegraphics[width=15cm]{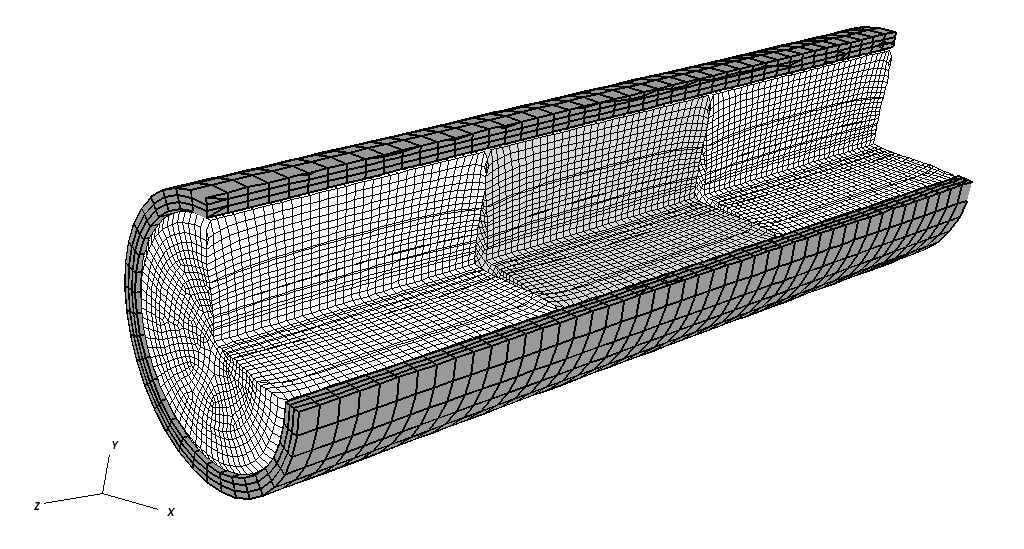}%
    \caption{A slice of a fuel rod mesh showing the outer clad mesh and three inner pellet meshes}
    \label{fig:pelletCladMeshes}%
\end{figure}

\begin{figure}%
    \centering
    \begin{subfigure}[b]{0.45\textwidth}
     \includegraphics[width=\textwidth]{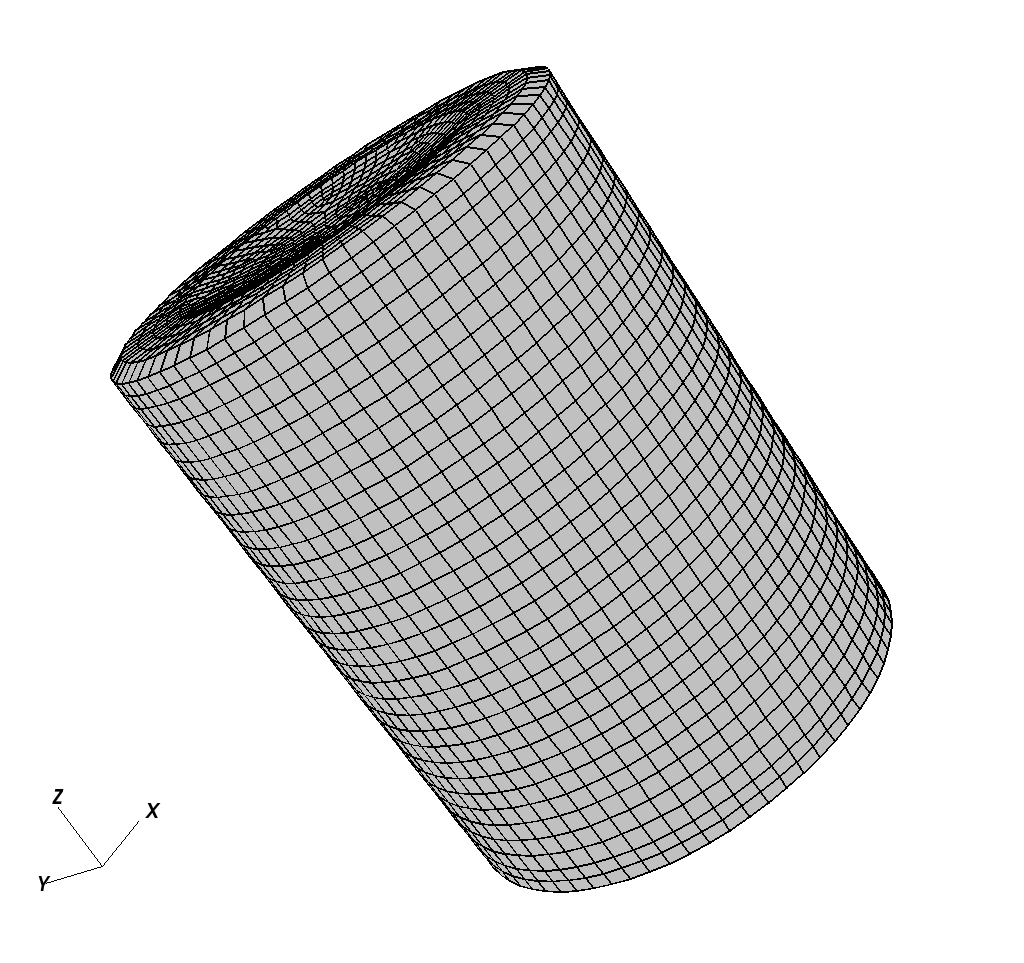}%
    \caption{IFA597 Pellet Mesh}
    \end{subfigure}
    \qquad
    \begin{subfigure}[b]{0.45\textwidth}
    \includegraphics[width=\textwidth]{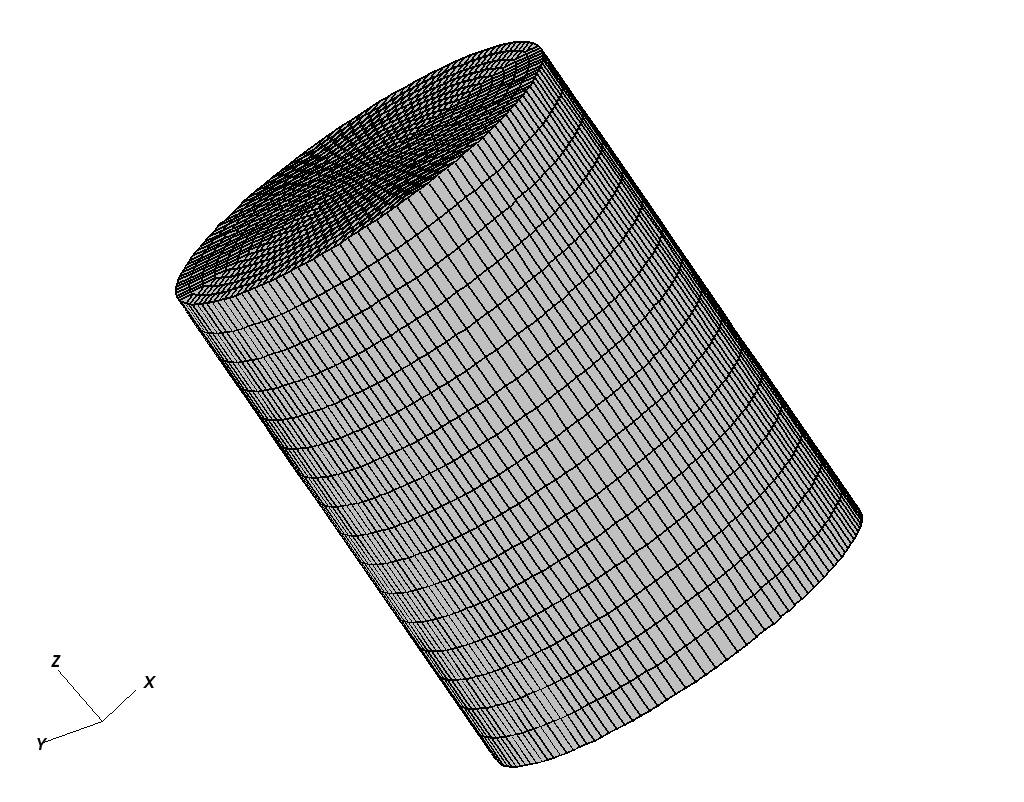}%
    \caption{IFA432 Pellet Mesh}
    \end{subfigure}
    \caption{(a) pellet mesh with dish and chamfer (b) standard pellet mesh }\label{pelletMeshes}
\end{figure}

As can be seen from Figures \ref{fig:pelletCladMeshes} and \ref{pelletMeshes}, the clad subdomain $\Omega_c$ and the pellet subdomains $\Omega_i$ are bounded, connected volumes with  piecewise smooth curved boundaries. For the purposes of this paper $\Omega_i$ is in general approximated by a polyhedral domain $\Omega_i^h\subsetneq \Omega_i$ during the mesh generation process. $\Omega_i^h$ is partitioned into a set, $\mathcal{P}_i^h$, of non-overlapping general hexahedral elements, $\mathcal{P}_i^h = \{\mathcal{P}_{i,1}^h, \mathcal{P}_{i,2}^h, \ldots, \mathcal{P}_{i,M}^h\}$ which are geometrically conforming, i.e.:
\begin{itemize}
\item $\mathop{\cup}\limits_{j=1}^M \mathcal{P}_{i,j}^h = \Omega_i^h$
\item The intersection of two elements, $\mathcal{P}_{i,l}^h$ and $\mathcal{P}_{i,m}^h$ for $l\neq m$ is either empty, a single vertex, an entire edge, or an entire face of both elements.
\end{itemize}

\subsection{Problem Discretization:}
A discrete approximation to $T_i$, the solution to Eqn (\ref{PelletWeakFormulation}) over the $i$-th pellet subdomain is obtained by solving the problem in a finite dimensional subspace $V_i^h$
of $V_i$. Let $dim(V_i^h) = N_i$ and $\{\phi^{h}_{i,k}\}_{k=1}^{N_i}$ be a suitable basis for $V_i^h$. Then, letting $T_i^h = \sum\limits_{k=1}^{N_i} T^h_{i,k}\phi^h_{i,k}$ where $T^h_{i,k}$ are unknown coefficients, the discrete Galerkin problem for each pellet is:\\
{\bf Discrete Pellet Sub-Problems:} Find $T_i^h =  \sum\limits_{k=1}^{N_i} T^h_{i,k}\phi^h_{i,k} \in V_i^h$ such that $\forall v^h \in V_i^h$
\begin{equation}
\begin{multlined}
\shoveleft[-0.25cm]{\langle k_i\nabla{T_i^h}, \nabla v^h \rangle_i + (h_{i,i-1}T_i^h, v^h)_{\Gamma_{i,i-1}} + (h_{i,i+1}T_i^h, v^h)_{\Gamma_{i,i+1}} + (h_{i,c}T_i^h, v^h)_{\Gamma_{i,c}}}\\
\shoveleft[-0.7cm]{= \;\langle f_i^h,v^h \rangle_i + (h_{i,i-1}T_{i-1}^m, v^h)_{\Gamma_{i,i-1}}+(h_{i,i+1}T_{i+1}^m, v^h)_{\Gamma_{i,i+1}}+(h_{i,c}T_c^m, v^h)_{\Gamma_{i,c}}}
\end{multlined}
\label{PelletDiscreteFormulation}
\end{equation}

Similarly, by a suitable choice of $V_c^h\subset V_c$ with $\text{span}\{\phi_{c,1}^h, \ldots, \phi_{c,N_c}^h\} = V_c^h$ and $dim(V_c^h) = N_c$ a discrete approximation to the solution of the clad problem (\ref{CladWeakFormulation}) can be obtained by solving:\\
{\bf Discrete Clad Sub-Problem:} Find $T_c^h =  \sum\limits_{k=1}^{N_c} T_{c,k}^h\phi_{c,k}^h \in V_c^h$ such that $\forall v^h \in V_c^h$
\begin{equation}
\begin{multlined}
\shoveleft[-3.0cm]{\langle k_c\nabla{T_c^h}, \nabla v^h \rangle_c + (h_{c,g}T_c^h, v^h)_{\Gamma_{c,g}} + (h_{c,f}T_c^h, v^h)_{\Gamma_{c,f}}}\\
\shoveleft[-2.0cm]{= \;\langle f_c^h,v^h \rangle_c + (h_{c,g}T_{g}^m, v^h)_{\Gamma_{c,g}}+(h_{c,f}T_{f}^m, v^h)_{\Gamma_{c,f}}}
\end{multlined}
\label{CladDiscreteFormulation}
\end{equation}

\noindent{\it Basis functions:} The use of general hexahedral or brick elements in discretizing the pellet and clad geometries leads us to use isoparametric finite elements for problem discretization. The basis functions over each geometric element $\mathcal{P}_{i,k}$ of a partition $\mathcal{P}_i$ are formed based on mappings of the standard trilinear $\mathbf{Q}_1$ basis functions defined over the reference cube $[-1,1]^3$ to $\mathcal{P}_{i,k}$. This ensures continuity across element boundaries within a partition. \\

All of the integral terms in Eqns ~(\ref{PelletDiscreteFormulation}) and ~(\ref{CladDiscreteFormulation}) are evaluated using a second order Gaussian quadrature.
For example, the term from Eqn ~(\ref{PelletDiscreteFormulation}) corresponding to the external source term, $\langle f_i^h,v^h \rangle_i$,
is computed by evaluating $f_i(\mathbf{x})$ at the quadrature nodes used by the finite element bases and
using the quadrature to evaluate the integrals, i.e.
\begin{equation}
\langle f_i^h,v^h \rangle_i = \sum_{j=1}^{M} \sum_{\ell=1}^{N_\ell} w_\ell \, f_i(\mathbf{x}_\ell) \, v^h(\mathbf{x}_\ell) \:,
\end{equation}
where the $j$ index denotes the spatial elements within the domain $\Omega_i^h$,
$N_\ell$ indicates the number of quadrature nodes per spatial element, and
$\mathbf{x}_\ell$ and $w_\ell$ are the quadrature nodes and weights, respectively.
Evaluating the source term at quadrature points rather than mesh nodes allows strict
conservation of the source term to be enforced automatically.
Because the source term is due to nuclear fission in the fuel pellets, the corresponding term
in Eqn~(\ref{CladDiscreteFormulation}) representing the source in the clad is treated as zero,
i.e. $\langle f_c^h,v^h \rangle_c = 0$.\\

For fixed $T_{i-1}^m,T_{i+1}^m$, and $T_{c}^m$, by choosing $v^h = \phi^{h}_{i,j}$ for  $j=1, \ldots, N_i$ in Eqn (\ref{PelletDiscreteFormulation}) we obtain a set of $N_i$ equations for the unknown coefficients $T^h_{i,k}, k=1, \ldots, N_i$ denoted by:
\begin{equation}
F_i(T_i^h) = 0,
\label{PelletNLE}
\end{equation}
where $T_i^h = (T_{i,1}^h, T_{i,2}^h, \ldots, T_{i,N_i}^h)^t$ with some abuse of notation. Since the conductivity $k_i$ and the effective heat coefficients $h_{i,i-1}$, $h_{i,i+1}$, and $h_{i,c}$ are nonlinear functions, Eqn (\ref{PelletNLE}) represents a coupled system of nonlinear algebraic equations over each pellet subdomain coupled through nonlinear boundary conditions to the adjacent pellets and clad. For the clad, a similar methodology leads to a set of $N_c$ nonlinear algebraic equations for the unknown coefficients $T^h_{c,k}, k=1, \ldots, N_c$ which we denote by:
\begin{equation}
F_c(T_c^h) = 0.
\label{CladNLE}
\end{equation}
The full set of nonlinear equations we wish to solve is given by 
\begin{equation}
\mathbf{F}(\mathbf{T})  = \mathbf{0}
\label{eqn::FullNLE}
\end{equation}
where $\mathbf{F}(\mathbf{T}) = \left( F_1(T_1^h), F_2(T_2^h), \ldots, F_N(T^h_N), F_c(T^h_c) \right)^t$ is the coupled set of block nonlinear equations across all pellet and clad domains with each block $F_i$ specified by Eqn (\ref{PelletNLE}) and $F_c$ given by Eqn (\ref{CladNLE}). Here, $\mathbf{T}$ denotes  the block unknowns across all domains, $\mathbf{T} \equiv \left( T^h_1, T^h_2, \ldots, T^h_N, T^h_c\right)^t$ with each $T_i$ a vector of unknowns itself.

\section{Solution Strategy}
\label{sec:algo}

Several nonlinear solution strategies could be used to solve the nonlinear system of algebraic equations described
in Section \ref{sec:discretization}. The fact that our problems often involve several hundred subproblems each discretized over a separate physical domain make solution methods that do not require formation of the full Jacobian matrix attractive. In particular we choose to use a Jacobian-Free Newton Krylov (JFNK) method \cite{knoll-jcp-04} for our nonlinear solver. The efficiency of a JFNK method when applied to a particular problem depends heavily on the preconditioner used. We will describe a physics based preconditioner that exploits the natural subdomain decomposition present in our application. A brief overview of the JFNK method is given in Section \ref{sec:jfnk} and the construction of the preconditioner is described in Section \ref{sec:pc}.

\subsection{Jacobian-free Newton-Krylov Methods}
\label{sec:jfnk}
Let $\mathbf{T}^{\star}$ denote the exact solution to Eqn (\refeq{eqn::FullNLE}). 
Classical Newton's method for solving Eqn (\refeq{eqn::FullNLE})
generates a sequence of approximations ${\mathbf{T}^k}$ to $\mathbf{T}^{\star}$, where
$\mathbf{T}^{k+1}=\mathbf{T}^k+\mathbf{s}^k$ and the Newton step $\mathbf{s}^k$ is the solution to
the system of linear equations 

\begin{equation}
\mathbf{J}^k\mathbf{s}^k=-\mathbf{F}(\mathbf{T}^k),
\label{eqn::Newtoneqns}
\end{equation} 
where $\mathbf{J}^k \equiv \mathbf{F}'(\mathbf{T}^k)$ is the Jacobian of $\mathbf{F}$ evaluated at $\mathbf{T}^k$. Newton's method is attractive because of its fast local convergence properties. 

For large-scale problems, Eqn (\refeq{eqn::Newtoneqns}) is typically solved using an iterative method because direct methods become impractical.  
Furthermore, it is often unnecessary to use a tight convergence tolerance for
 the iterative method when $\mathbf{T}^k$ is far from $\mathbf{T}^{\star}$, since the linearization that leads to 
(\refeq{eqn::Newtoneqns}) may be a poor approximation to $\mathbf{F}(\mathbf{T})$. 

Generally, it is much more efficient to employ inexact Newton methods \cite{DemboEisentatSteihaug82}, 
in which the convergence tolerance for (\ref{eqn::Newtoneqns}) is selected adaptively by requiring that
$\mathbf{s}^k$ only satisfy: 

\begin{equation}
\norm{\mathbf{F}(\mathbf{T}^k)+\mathbf{J}^ks^k}\leq\eta_{k}\norm{\mathbf{F}(\mathbf{T}^k)}
\label{eqn::inexactNewtoneqns}
\end{equation}
for some $\eta_{k}\in(0,1)$ \cite{DemboEisentatSteihaug82}. With an appropriate choice of the forcing term $\eta_{k}$ superlinear and even quadratic convergence of the iteration can be achieved \cite{EisenstatWalker94}.

While any iterative method can be used to find an $\mathbf{s}^k$ that satisfies (\refeq{eqn::inexactNewtoneqns}), Krylov 
subspace methods are distinguished by the fact that they only require matrix-vector products to proceed. These
 matrix-vector products can be approximated by a finite-difference version of the directional (G\^ateaux) derivative as:

\begin{equation}
 \mathbf{J}^k\mathbf{v}\approx\frac{\mathbf{F}(\mathbf{T}^k+\varepsilon \mathbf{v})-\mathbf{F}(\mathbf{T}^k)}{\varepsilon},
\label{eq:Jac-free}
\end{equation} 
which is especially advantageous when $\mathbf{J}^k$ is difficult to compute or expensive to store (as is the case in this 
application due to the presence of multiple meshes). While the selection of a suitable differencing
parameter $\varepsilon$ may be non-trivial for some applications, it
is generally well-understood \cite{kelley}. For this application, we choose:
\[\varepsilon = \sqrt{\epsilon_{\mathrm{mach}}} \frac{\sqrt{1+\|\mathbf{T}^k\|}}{\|\mathbf{v}\|},\] 
where $\epsilon_{\mathrm{mach}}$ is machine precision and $\|\cdot\|$
refers to the $l_2$-norm. In our applications, which are performed in
double precision, $\varepsilon$ is typically on the order of
$10^{-10}$.

From the various Krylov methods available, GMRES was selected
because it guarantees convergence with nonsymmetric, 
nonpositive definite systems \cite{saad-gmres} (the case in some of our examples), and because it provides normalized 
Krylov vectors $\|\mathbf{v}\|=1$, thus bounding the error introduced in the difference approximation of (\ref{eq:Jac-free}) (whose
 leading error term is proportional to $\varepsilon\| \mathbf{v}\|^{2}$) \cite{mchugh94}.
 
 However, GMRES can be memory intensive (storage increases linearly with the number of GMRES iterations per Jacobian solve) and 
 expensive (computational complexity of GMRES increases quadratically with the number of GMRES iterations per
 Jacobian solve). In principle, restarted GMRES can deal with these limitations; however, it lacks a theory of convergence,
 and stalling is frequently observed in real applications \cite{knoll-siam-1998}. In our applications, we rely on performing
 inexact solves combined with efficient preconditioning to keep the number of GMRES iterations required to compute each 
 inexact Newton step, $\mathbf{s}^k$, small.

\subsection{Preconditioning}
\label{sec:pc} Preconditioning is a numerical technique used within an iterative method to accelerate the process of finding
 a solution to a system of equations. The use of a preconditioner will increase the cost of each iteration but
 a good preconditioner will drastically reduce the total number of iterations required to solve the system of 
 equations thus making the overall process significantly faster than the unpreconditioned case. While
 the idea can be applied to both nonlinear and linear systems of equations, we only focus on applying to the linear systems at each Newton step
within the JFNK procedure. In particular, we used a ``right preconditioning'' 
 procedure to solve (\ref{eqn::Newtoneqns}). In this approach, (\ref{eqn::Newtoneqns}) is transformed to the 
 equivalent system shown in (\ref{eqn:rightPC}) by using a preconditioner, $M$. 
  
\begin{equation}
\left(\mathbf{J}\mathbf{M}^{-1}\right)\mathbf{M} \mathbf{s} = -\mathbf{F}(\mathbf{T})
\label{eqn:rightPC}
\end{equation}  

The above system is solved in two steps: (1) solve $\mathbf{A} \mathbf{y} = -\mathbf{F}(\mathbf{T})$ using the GMRES method where $\mathbf{A} = \mathbf{J}\mathbf{M}^{-1}$
 and (2) compute $\mathbf{s} = \mathbf{M}^{-1} \mathbf{y}$. The matrix-vector product, $\mathbf{A} \mathbf{v}$, is approximated as shown in (\ref{eqn:pcjacMatVec}).  

\begin{equation}
\mathbf{A} \mathbf{v} \approx \frac{\mathbf{F}(\mathbf{T} + \varepsilon \mathbf{M}^{-1} \mathbf{v}) - \mathbf{F}(\mathbf{T})}{\varepsilon},
\label{eqn:pcjacMatVec}
\end{equation}

Ideally, $\mathbf{M}^{-1}$ should be a good approximation to $\mathbf{J}^{-1}$, it should be easy and inexpensive to compute and apply, and 
the computation and application of $\mathbf{M}^{-1}$ should have good parallel scalability. It is difficult to meet all of these competing
requirements and some trade-offs need to be made while designing $\mathbf{M}$.
 
 The design of the preconditioner used for our problem stems from a careful observation of the structure of the true Jacobian matrix which is of the form:
\begin{equation}
 \mathbf{J} = \begin{bmatrix}
J_{11} & J_{12} & 0          & 0         & \cdots  & 0         & J_{1C} \\
J_{21} & J_{22} & J_{23} & 0         & \cdots & 0         & J_{2C}\\
0         & J_{32} & J_{33} & J_{34} & \cdots & 0         & J_{3C}\\
0         & 0         & J_{43} & J_{44} & \cdots & 0         & J_{4C} \\
\vdots & \vdots & \vdots   & \vdots & \ddots & \vdots & \vdots \\
0         & 0         & \cdots   & \cdots & \cdots  & J_{NN}        & J_{NC}\\
J_{C1}&J_{C2} & \cdots   & \cdots & \cdots  & J_{CN} & J_{CC}\\
\end{bmatrix}
 \end{equation}
 Here, each $J_{ii}, \; i=1, \ldots, N$ is a block matrix denoting the portion of the full Jacobian arising from interior-interior connections within the $i$-th pellet.
 $J_{i,i-1}$ and $J_{i,i+1}$ are block matrices corresponding to the boundary interactions between pellet $i$ and pellets $(i-1)$ and $(i+1)$ respectively.  $J_{iC}$ and $J_{Ci}$
 are block matrices corresponding to the boundary couplings between the $i$-th pellet and the clad.
 
 Under normal operating conditions within a reactor the heat transfer between adjacent pellet domains is considerably lower than between the pellets and clad. This prompts us to drop off diagonal terms corresponding to pellet-pellet interactions when forming the preconditioning matrix $\mathbf{M}$. This leaves off diagonal coupling terms between the pellets and clad. While a preconditioner that retains these terms is likely to yield better overall convergence it requires a block Gauss-Seidel type approach that limits the level of asynchrony within the preconditioner. By dropping off diagonal terms corresponding to pellet-clad interactions we choose to sacrifice potentially better convergence for the ability to asynchronously solve all domains at once within the preconditioner solve step. Finally, block diagonal terms of the Jacobian that involve the partial
 derivative of the thermal conductivity, $k$, with respect to the temperature, $T$ are ignored to yield an approximate Jacobian of the form
 \begin{equation}
 \mathbf{M} = \begin{bmatrix}
\tilde{J}_{11} & 0 & 0          & 0         & \cdots  & 0         & 0 \\
0 & \tilde{J}_{22} & 0 & 0         & \cdots & 0         & 0 \\
0         & 0 & \tilde{J}_{33} & 0 & \cdots & 0         & 0 \\
0         & 0         & 0 & \tilde{J}_{44} & \cdots & 0         & 0 \\
\vdots & \vdots & \vdots   & \vdots & \ddots & \vdots & \vdots \\
0         & 0         & \cdots   & \cdots & \cdots  & \tilde{J}_{NN}        & 0 \\
0         & 0         & \cdots   & \cdots & \cdots  & 0 & \tilde{J}_{CC} \\
\end{bmatrix}
\label{PCSystem}
 \end{equation}
Inverting systems of the form $\mathbf{M}\mathbf{s} = \mathbf{y}$ now correspond to solving $(N+1)$ independent subsystems
\begin{equation}
\tilde{J}_{ii} s_i = y_i
\label{PCSubSystems}
\end{equation}
At the $k$-th Newton step, systems $1$ through $N$ correspond to discretizing and solving $N$ variable coefficient {\it linear}  diffusion PDE systems, one for each pellet domain, of the form:
\begin{equation}
-\nabla\cdot k_i(T^k_i, \mathbf{x})\nabla{T_i}(\mathbf{x}) = r^k_i(\mathbf{x}), \; \text{for} \; \mathbf{x}\in \Omega_i,
\label{LinearPelletThermalTransport}
\end{equation}
with boundary conditions
\begin{eqnarray}
k_i(T^k_i)\nabla T_{i}\cdot \mathbf{n}_i + h_{i,i-1}(T^k_i, 0) T_{i}  &=& 0 \; \text{on} \;\Gamma_{i,i-1}, \label{P2PLinearRobinBC1}\\
k_i(T^k_i)\nabla T_{i}\cdot \mathbf{n}_i + h_{i,i+1}(T^k_i, 0) T_{i} &=& 0 \; \text{on} \;\Gamma_{i,i+1}.
\label{P2PLinearRobinBC2}
\end{eqnarray}
where $T^k_i$ is the current approximation to the solution of the nonlinear system (\ref{eqn::FullNLE}) over the $i$-th pellet domain and $r^k_i$ is the nonlinear residual.
System $(N+1)$ of the preconditioner solve is over the clad domain and again involves discretizing and solving a {\it linear} variable coefficient diffusion PDE system of the form:
\begin{equation}
-\nabla\cdot(k_c(T^k_c, \mathbf{x})\nabla{T_c}) = r^k_c \;\text{for}\; \mathbf{x} \in \Omega_c
\label{LinearCladThermalTransport}
\end{equation}
with boundary conditions
\begin{eqnarray}
k_c(T^k_c)\nabla T_{c}\cdot \mathbf{n_g} + h_{c,g}(T^k_c, 0) T_{c}  &=& 0 \; \text{on} \;\Gamma_{c,g} \\
k_c(T^k_c)\nabla T_{c}\cdot \mathbf{n_f} + h_{c,f}(T^k_c, 0)T_{c}  &=& 0 \; \text{on} \;\Gamma_{c,f}
\label{LinearCladRobinBC}
\end{eqnarray}
with $T^k_c$ being the current approximation to the solution of the nonlinear system (\ref{eqn::FullNLE}) over the clad domain and $r^k_c$ the nonlinear residual. Discretization of the linear variable coefficient diffusion PDE systems
described above follows along the lines of the methodology described in section (\ref{sec:discretization}). Each application of the preconditioner involves $(N+1)$ algebraic multigrid solves to invert the systems
 (\ref{PCSubSystems}). Since these are independent subsystems all of the $(N+1)$ solvers can operate simultaneously when distributed over different processor sets enabling us to obtain a high degree of parallelism irrespective of the number of pellet subdomains present.

%
%
%
%
%
%
%
%

\section{Computational Infrastructure}
\label{sec:infrastructure}
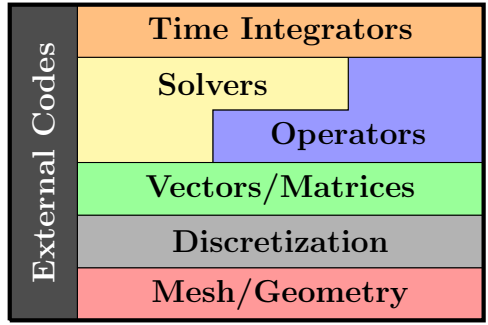
\begin{figure}[ht]
\begin{center}
\begin{tikzpicture}[node distance = 0, auto, xscale=0.9, yscale=0.7]]
    \draw [black, fill=black!70]  (0,0) -- (1,0) -- (1,6) -- (0,6) -- (0,0);
    \draw [black, fill=red!40]    (1,0) -- (7,0) -- (7,1) -- (1,1) -- (1,0);
    \draw [black, fill=black!30]  (1,1) -- (7,1) -- (7,2) -- (1,2) -- (1,1);
    \draw [black, fill=green!40]  (1,2) -- (7,2) -- (7,3) -- (1,3) -- (1,2);
    \draw [black, fill=blue!40]   (3,3) -- (7,3) -- (7,5) -- (5,5) -- (5,4) -- (3,4) -- (3,3);
    \draw [black, fill=yellow!40] (1,3) -- (3,3) -- (3,4) -- (5,4) -- (5,5) -- (1,5) -- (1,3);
    \draw [black, fill=orange!50] (1,5) -- (7,5) -- (7,6) -- (1,6) -- (1,5);
    \draw [ultra thick]          (0,0) -- (7,0) -- (7,6) -- (0,6) -- (0,0);
    \node [text=white, rotate=90] at (0.5,3.0) {\textbf{External Codes}};
    \node [] at (4,5.5) {\textbf{Time Integrators}};
    \node [] at (3,4.5) {\textbf{Solvers}};
    \node [] at (5,3.5) {\textbf{Operators}};
    \node [] at (4,2.5) {\textbf{Vectors/Matrices}};
    \node [] at (4,1.5) {\textbf{Discretization}};
    \node [] at (4,0.5) {\textbf{Mesh/Geometry}};
\end{tikzpicture}
\caption{Structure of components in AMP}\label{fig:arch}
\end{center}
\end{figure}

Several computational tools are necessary for performing simulations such as the one described in this paper; we developed the 
Advanced Multi-Physics (AMP) \cite{ampTM2011} package for this purpose. AMP is a complete system for simulating stationary and time 
dependent, multi-domain, coupled physics problems. AMP consists of several software components. Each 
component is designed to provide a uniform consistent interface which interacts with other components, and developers 
of other components are only exposed to these interfaces. This is despite the fact that AMP is designed to sit in between
existing software frameworks to leverage their strengths and investments without over-dependence. The complexities of 
interfacing with different software frameworks are kept behind the standard interfaces that AMP provides. Fig. \ref{fig:arch}
illustrates the structure of the various components in AMP. A brief description of each of these components is given below.

\subsection{Mesh and geometry}
The mesh and geometry interface (AMP::Mesh) allows AMP to interact 
with multiple mesh or geometry packages. AMP::Mesh already interfaces with the 
LibMesh \cite{libMeshPaper} and STKMesh \cite{stkmesh:website} packages, in addition to maintaining a native structured mesh 
capability.  The native structured mesh capability includes automated mesh generation for
simple geometries enabling users to create boxes, cylinders, and tubes without an external
mesh generation tool.  These internal mesh generators were used for most of the results 
presented in section \ref{sec:results}.

\subsection{Discretization}
Due to the close coupling between mesh and discretization, AMP::Mesh (using functionality 
contained in LibMesh), currently handles the discretization also. This will become a separate 
interface, if the need arises for us to interface different discretization packages with mesh packages.

\subsection{Vectors and matrices}
AMP provides standard AMP::Vector and AMP::Matrix classes which serve two purposes. 
Firstly, they provide users with a standard interface to perform vector and matrix operations. 
At the same time, the classes hide the details of interfacing with various software packages 
that have their own definition of vectors and matrices. For example, Trilinos \cite{Trilinos-Overview} and PETSc \cite{petsc-home-page} both
provide matrix operations and Trilinos, PETSc, and SUNDIALS \cite{sundials:website} provide and/or use vector operations. 
AMP Vector and Matrix act as the interfaces to these packages through the Vector and Matrix classes and enable a user to combine components from all of these packages to build powerful AMP applications while not having to tackle the complexities of
interacting with each of these packages. 

The vector class also contains the ability to compose multiple parallel vectors into a single multi-vector
that can be provided to a solver.  For example, the global solution and source vectors are created
as a composition of nodal vectors on each mesh and composed into a single multi-vector.  A view to
the multi-vector is provided to PETSc SNES nonlinear solver for the Newton iterations and a view of the vector and the diffusion matrix 
associated with a single domain is provided to Trilinos ML for the multi-grid preconditioning.

\subsection{Operators}
Operators are the core of the AMP design and where all of the physics is contained. 
Operators encapsulate the details of the mapping operation $\mathcal{L}: X \rightarrow Y$ where 
$X$ and $Y$ are appropriately defined spaces. Operators may represent discretized PDE operators, 
boundary operators, an operation to extract material properties from material databases or tables, 
linear or nonlinear algebraic operations, or compositions of the above. The ability to compose 
operators and to extract information from compositions is intended to facilitate the incremental 
construction of multi-physics and/or multidomain simulations as well as rapid prototyping and 
experimentation to understand couplings in multi-physics simulations. The nonlinear and linear
FEM operators for diffusion, boundary conditions and interpolation maps between domains are
encoded as operators.

\subsection{Solvers} 
Solvers in AMP refer to the nonlinear and linear solvers that represent the action 
of an approximate inverse map of a given operator if that inverse operation has some well defined 
meaning. In this sense solvers can also be considered as operators.  
An inverse operator can be easily constructed by wrapping a solver in an inverse operator 
class. The solver interface allows the user to utilize a standard interface to solvers from 
Trilinos, PETSc, native AMP solvers, and potentially 
other packages in the future. Again, the design emphasis has been to provide a standard interface to 
hide the complexity of particular software packages from a user and to avoid  required dependence on a 
particular software package. The solver interface enables us to create the complex nonlinear solvers, linear solvers,
and preconditioners across multiple domains without significant code rewrites.

\subsection{Time integrators}
AMP time integrators provide a uniform interface to solving time-dependent systems 
which can include Differential Algebraic Equations (DAEs). This is necessary within the context 
of our broader target application class because of coupling between time dependent thermal and quasi-static 
mechanical systems being simulated. The design allows for explicit, semi-implicit, and fully 
implicit simulations of coupled multi-physics problems. In the case of semi-implicit and fully 
implicit calculations, the solver interfaces in AMP are used, and in all cases, 
the operator interfaces are used to allow composable multi-physics simulations allowing users 
to experiment with coupling different physics together. The time integrator interface is used 
to provide an interface to the SUNDIALS suite of time integrators and can be used 
in future to interface to other time integrator packages such as the Rhythmos package of Trilinos. 

\subsection{External packages}
AMP is designed to leverage existing software whenever possible including off-the-shelf leadership class computational packages that
include the Trilinos, PETSc and SUNDIALS packages.
In general the infrastructure design of AMP does not rely on any external software, but provides interfaces for
using external software within AMP.  For example, a user can create a vector using AMP's internal vector, a PETSc vector,
or a Trilinos Epetra vector, but can then use the given vector within the solvers that may use PETSc or Trilinos solvers.  
AMP leverages capabilities within many software packages
through a seamless application programming interface including
MPI for parallel capabilities, PETSc for vectors, nonlinear and linear solvers, Trilinos for vectors, nonlinear solvers, and algebraic multigrid solvers, SUNDIALS for implicit time integrators, LibMesh and STKMesh for discretization and meshing and HDF5 and Silo for IO.

\subsection{Parallel implementation}
AMP is primarily designed to be a parallel infrastructure based on MPI.  An MPI-based utility class is provided that allows the 
user to utilize MPI. The interface enables AMP to be compiled without MPI for users who do not wish to leverage the parallel capabilities.  The core design is independent of the parallelization, 
and the parallelization is based on parallel decomposition of the meshes.  There is a two-level parallel decomposition
used for the mesh domains.  First, the individual mesh domains are split onto separate communicators and independent
processor groups.  This minimizes the number of processors per mesh and ensures that independent meshes can 
utilize collective operations that do not include the processors of other meshes.  This splitting is done internally
within AMP and can be modified by the user.  A second level of domain decomposition is then performed to divide each
mesh domain among the processors in the MPI group for that domain.  This level of decomposition is handled by the 
package responsible for the current mesh domain.  For example, if the underlying mesh for a given domain is LibMesh, it
will perform the decomposition, while a native AMP mesh will be controlled by AMP.  Note that different mesh domains
may be owned by different packages and this is fully supported.  

Once the domain-decomposition is performed, vectors and matrices may be created over a single mesh, an arbitrary 
combination of meshes, or a subset of a mesh (or multiple meshes).  Each vector or matrix exists over a given 
communicator that does not need to match any mesh.  Linear operations are then performed on this communicator reducing
the need for global operations over the entire global communicator.  Maps between multiple domains use
a communicator that spans two or more existing communicators.  The MPI utility class provides all routines for 
creating and managing the communicators with MPI, including their proper destruction when they are no longer needed.
Section \ref{sec:strongScaling} includes the results of a scaling study conducted using the problem described in Section 
\ref{sec:testSetup} and using the parallel load balancing strategies described here.

\section{Numerical Experiments}
\label{sec:results}

A suite of numerical experiments were defined to verify the accuracy of the thermal transport capability of AMP for a multi-domain problem that is based on the geometry and materials of nuclear fuel.  Independent studies were performed to verify the accuracy of the solution using the method of manufactured solutions (Section \ref{sec:verification}), evaluate the accuracy of the code with respect to experimental data and a well characterized code used for regulatory analysis (Section \ref{sec:validation}), and evaluate the scalability of the parallel algorithm (Section \ref{sec:scaling}).  All of the studies were based on actual geometries and material properties for experimental nuclear fuel rods that are defined in Section \ref{sec:testSetup}. All numerical experiments were performed on the Titan (Cray XK7) and EOS (Cray XC30) supercomputers hosted at the Oak Ridge Leadership Computing Facility with 8 and 16 MPI processes per compute node respectively. Load balancing is done automatically by the load balancer within AMP. The AMP nonlinear solver internally leveraged the JFNK implementation within the PETSc package with absolute and relative tolerances for the JFNK nonlinear solver being set to 1.0e-12 and 1.0e-10 respectively. Right preconditioned FGMRES with a maximum Krylov dimension of $40$ was used within our simulations. The AMP preconditioner consists of a block Jacobi solver as described earlier with each block Jacobi solver component consisting of one or more iterations of an algebraic multigrid (Trilinos ML) V-cycle solver with 2 pre- and post-smoothing steps of a symmetric Gauss-Seidel smoother, a maximum of 10 multigrid levels and a coarse grid direct solver. All simulations were performed in double precision arithmetic.

\subsection{Experimental Setup}
\label{sec:testSetup}

The materials and geometries used in the following numerical experiments are based on one of two well-characterized experiments from the International Fuel Performance Experiments (IFPE) database \cite{ifpe:website}.  

\begin{figure}
\centering
    \includegraphics[width=\textwidth]{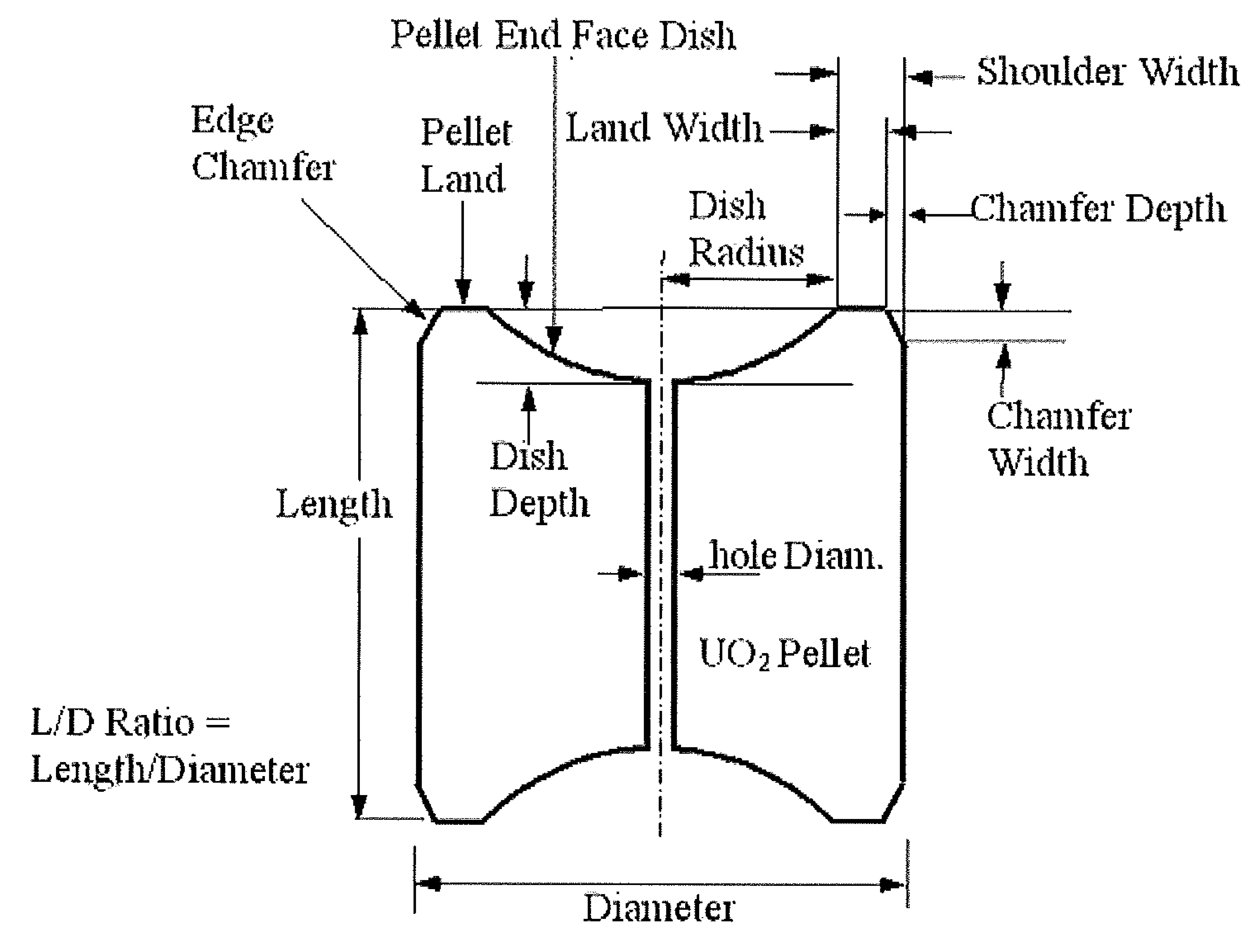}
    \caption{Mid-plane slice of an annular nuclear fuel pellet with a dish and chamfer}
    \label{fig:complexPellet}
\end{figure}

The Integrated Fuel Assembly (IFA) 432 \cite{ifa432:website}, Rod 1, is a standard nuclear fuel rod (uranium-dioxide or, UO$_2$, fuel in Zircaloy-4 clad) that was irradiated in the Halden boiling water reactor from December 1975 to June 1982 with online temperature measurements at one axial location in the center of the fuel.  
The IFA 597 \cite{ifa597:website}, Rod 2, contains weapons-grade mixed-oxide (MOX) fuel within Zircaloy-4 clad in a more modern geometry that includes a dish, chamfer, and central hole, as shown in Figure \ref{fig:complexPellet}.
IFA 597, Rod 2 was irradiated in the Halden boiling water reactor from from July 1997 to January 2002 with online temperature measurements at one axial location in the center fuel.  Each of these rods is a short version of a full length commercial fuel rod.
The geometric and material description of the experimental rods are provided in Table \ref{tab:valGeometry} for the IFA 432 and IFA 597 experiments.  
The clad height and number of pellets modeled were adjusted in Sections \ref{sec:verification} and \ref{sec:scaling} to achieve the purpose of that section.  However, these geometries were used exactly in Section \ref{sec:validation}.
The thermal conductivity ($k$) of Zircaloy-4 (Eq. \ref{eq:kzr}) and fuel (Eq. \ref{eq:kfuel}), including both  UO$_2$ and MOX, depend on the temperature (T) and burnup (B), which is a measure of total heat generated locally.

\begin{table}[htbp]
   \centering
   \begin{tabular}{@{} |lr|c|c| @{}} 
      \hline
      Dimension & Units & IFA 432 & IFA 597 \\
      \hline
      Pellet ID & mm & 0 & 1.8  \\      
      Pellet OD & mm & 10.67 & 8.05 \\
      Pellet Dish Depth & mm & 0 & 0.26 \\
      Pellet Dish Diameter & mm & 0 &  2.15 \\      
      Pellet Chamfer Height & mm & 0 & 0.15 \\
      Pellet Chamfer Diameter & mm & 0 & 5.3 \\      
      Pellet Density & g/cc & 10.42 & 10.5 \\      
      Pellet Height & mm & 13 & 10.5  \\      
      Number of Pellets & & 44 & 21 \\
      Clad ID & mm & 10.9 & 8.22 \\      
      Clad OD & mm & 12.78 & 9.5  \\
      Clad Height & mm & 622.8 & 252.\\
      \hline
   \end{tabular}
   \caption{Geometry and material specification for validation problems}
   \label{tab:valGeometry}
\end{table}

\begin{equation}
  k[\text{Zr}] = 7.51 + 2.09 \times 10^{-2} T  -1.45 \times 10^{-5} T^2 + 7.67 \times 10^{-9} T^3 \label{eq:kzr}
\end{equation}

\begin{equation}
k[\text{Fuel}] = \alpha \left ( \left ( \beta + \gamma T + \delta \left (1.0 +\epsilon e^{-\frac{\zeta}{T}} \right )^{-1} \right ) ^{-1} + \eta T^{-2} e^{-\frac{\theta}{T} } \right ) \label{eq:kfuel} 
\end{equation}

\begin{eqnarray}
\text{where} \quad \alpha[\text{UO}_2] =& 1.00767; \: & \alpha[\text{MOX}] = 1.05353 \nonumber \\
\beta[\text{UO}_2] =&  0.0452 + 0.00187 B; \: & \beta[\text{MOX}] = 0.035 + 0.00187 B \nonumber \\
\gamma[\text{UO}_2] =&  0.000246;  \: & \gamma[\text{MOX}] = 0.000286  \nonumber\\
\theta[\text{UO}_2]  =& 16361; \: & \theta[\text{MOX}]  =  13520 \nonumber
\end{eqnarray}
\begin{eqnarray}
\delta &=&  0.038  \left (1.0 - 0.9 e^{-0.04 B} \right ) B^{0.28}  \nonumber \\
\epsilon &=&  396 \nonumber \\
\zeta &=&  6380 \nonumber \\
\eta &=&  3.5  \times 10^9 \nonumber 
\end{eqnarray}

To define the heat source in the nuclear fuel, AMP allows the user to either define the power distribution, $f(r, \theta, z)$, as a function of the radius ($r$) from the center of the pellet, height ($z$), and azimuthal-angle ($\theta$) about the $z$-axis or provide a power distribution in a coupled-physics calculation at every quadrature-point in the problem.
The user-defined power definition allows for a simple definition of the axial shape functions or complex nuclear-specific features, such as the radial rim effect or azimuthal variations guide tubes and control rods.
The radial power shape includes the option to use a model that coincides with the empirically-derived TUBRNP model (Equation \ref{eq:frap}) from the Transuranus nuclear fuel performance code \citep{transuranus_08}.
\begin{equation}  
\label{eq:Cyl}
f(r, \theta, z) = 1+ a\: F(r) + b_{\theta}sin(\theta) + \sum_{k>0}{c_k P_k(z)}; \\
\end{equation}
\begin{equation}  
\label{eq:frap}
F(r) = 1+ 3.45 \: exp[-3(R-r)^{0.45}].
\end{equation}

In equation \ref{eq:Cyl}, the user-defined coefficients ($a$, $b_{\theta}$, and $c_k$) define the magnitude of each component, $P_k(z)$ are Legendre polynomials, and $F(r)$ is the TUBRNP model that is based on the radius and the outer radius of the fuel pellet ($R$).
In the verification testing, a manufactured-source was utilized; the validation testing utilized the TUBRNP model.

\subsection{Verification Studies}
\label{sec:verification}

\begin{figure}
   \centering
   \includegraphics[scale=0.5]{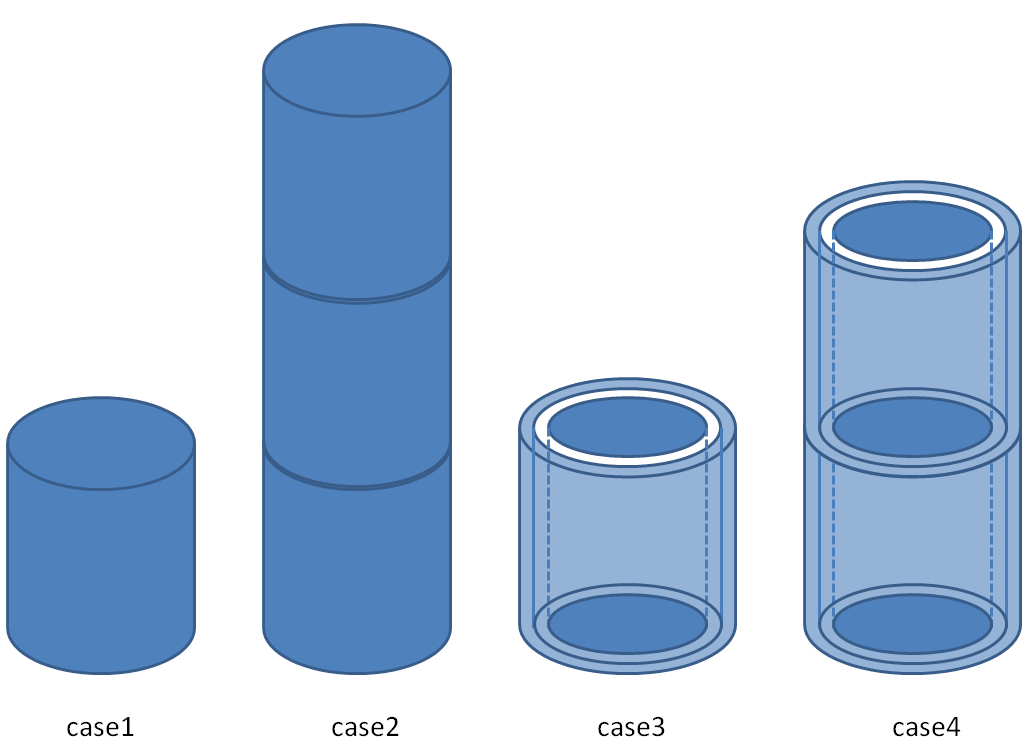}
   \caption{Schematic of pellet and clad geometries used in verification studies} 
   \label{fig:verification-schematic}
\end{figure}

Verification studies for modeling steady state thermal contact for nuclear fuels are presented here. The verification process uses the method of manufactured solutions. For this study, the pellet geometry is based on the IFA 432 experiment and the clad geometry is simplified as shown in Figure \ref{fig:verification-schematic}.  Material properties of UO$_2$ for the pellet and Zircaloy for the clad are based on the IFA 432 experiment as listed in Section \ref{sec:testSetup}. The pellet and clad domains are not in contact and have a (gap) distance between the surfaces. A total of four cases are studied using this configuration. \\

\noindent {\bf{Case 1}}: Three-dimensional single fuel pellet (with no clad) that exhibits strong gradients in all directions with different Robin boundary conditions between circular surface along the height and end surfaces.\\ 
\noindent {\bf{Case 2}}: Three fuel pellets stacked upon each other (with no clad). The surfaces are in contact and a manufactured solution is constructed using the fuel pellet contact conductance model. At the contact surfaces a Robin boundary condition is imposed and the results would make apparent any anomalies in the volume and boundary condition discretizations. \\
\noindent {\bf{Case 3}}: Single fuel pellet and clad with heat transfer across a gap between the surfaces. The manufactured solution for both the domains is constructed taking the nonlinear gap conductance between the internal surfaces into consideration. The results would make apparent any lack of energy conservation due to heat transfer across the gap.\\
\noindent {\bf{Case 4}}: Two pellets and clad including contact between pellets and heat transfer across a gap between pellet and clad.  This verifies the implementation at corner points which intersect multiple domains. \\

\noindent We begin by selecting the exact solution to be
\begin{equation}
\label{manufacturedSol}
\phi(x,y,z) = 800 z + 10^6(0.00004-20 x^2-20 y^2)
\end{equation}
which is qualitatively similar to the thermal solutions and in bounds to the material models. Substituting it for $T$ in the differential equation (\ref{PelletThermalTransport}) we obtain the corresponding analytic right hand side. These sources and corresponding boundary conditions are evaluated at each quadrature point of the entire finite element mesh in order to eliminate interpolation errors. The relative and absolute convergence tolerances within the nonlinear solver are set to 1.0e-10 to ensure solver errors are below discretization error bounds. The discretization error is evaluated using an $L_2$ norm.
\begin{equation}
||\phi-\phi_{h}||_{L_2} = \left(\int_{\Omega}{(\phi-\phi_{h})^{2}d\Omega}\right)^{1/2} \approx \left(\sum_{\Omega}\sum_{qp}(\phi-\phi_{h})^{2} J(x_{qp})w_{qp}\right)^{1/2}
\end{equation}
Convergence rates of these norms are reported on progressively refined meshes by increasing the number of elements in all three cylindrical coordinates to simplify computing the characteristic element length $``h"$ of the unstructured mesh. Also, we make sure that the refinements are generated from the original geometry to ensure that the meshes are geometrically conforming. The rate of convergence is given by 
\begin{equation}
p=\frac{log\left(\frac{||e_2||}{||e_1||}\right)}{log\left(\frac{h_2^2}{h_1^2}\right)} \approx \frac{log\left(\frac{||e_2||}{||e_1||}\right)}{log\left(4\right)}
\end{equation}

\begin{table}
\centering
\begin{tabular}{|lllll|}
\hline
Problem & \# Elements & $||\phi-\phi_{h}||_{L_\infty}$ & $||\phi-\phi_{h}||_{L_{2}}$ & $p$\\
\hline
\multirow{3}{*}{ {\bf{Case 1}}} &1890 & 2.1796 & 0.002411& - \\
                                &14944 &0.7426 & 0.000624 & 1.94\\
                                &119296 &0.2256 & 0.000155 & 2.00\\
\hline
\multirow{3}{*}{ {\bf{Case 2}}} & 5670 & 2.1845 & 0.004139& - \\
                                & 44832 & 0.7810 & 0.001073& 1.94\\
                                & 357888 & 0.2644 & 0.000267& 2.00\\
\hline
\multirow{3}{*}{ {\bf{Case 3}}} & 3510  & 2.9740 & 0.003132& - \\
                                & 27904 & 0.7053 & 0.000804 & 1.96\\     
                                & 222976& 0.1817 & 0.000200 & 2.00\\
\hline
\multirow{3}{*}{ {\bf{Case 4}}} & 5400 & 2.9851 & 0.004434 & - \\
                                & 42848& 0.7446 & 0.001139& 1.96\\
                                & 342272& 0.2207 & 0.000284& 2.00\\
\hline
\end{tabular}
\caption{Mesh refinement studies}
\label{tab:meshRefinement}
\end{table}

\begin{table}[H]
\centering
\begin{tabular}{|llll|}
\hline
\# Elements & $||\phi-\phi_{h}||_{L_\infty}$ & $||\phi-\phi_{h}||_{L_{2}}$ & $p$\\
\hline
{\bf{10 Pellets}}&&&\\
18900 & 2.1848 & 0.007606& \\
149440 & 0.7810 & 0.001971& 1.94\\
1192960 & 0.2644 & 0.000491& 2.00\\
\hline
{\bf{50 Pellets}}&&&\\
94500 & 2.1223 & 0.017045& \\
756000 & 0.7810 & 0.004416& 1.94\\
6048000 & 0.2644 & 0.001100& 2.00\\
\hline
\end{tabular}
\caption{Many domain mesh refinement studies for Case 2}
\label{tab:meshRefinement2}
\end{table}

\begin{table}[H]
\centering
\begin{tabular}{|llll|}
\hline
Elements & $||\phi-\phi_{h}||_{L_\infty}$ & $||\phi-\phi_{h}||_{L_{2}}$ & $p$ \\
\hline
{\bf{10 Pellets}}&&&\\
35100 &  2.9899 & 0.009937& \\
280800&  0.7450& 0.002549& 1.96\\
2246400& 0.2210& 0.000636& 2.00\\
\hline
{\bf{50 Pellets}}&&&\\
87750 & 2.9852 & 0.022190& \\
702000 & 0.7450 & 0.005700& 1.96\\
5616000 & 0.2210 & 0.001421& 2.00\\
\hline
\end{tabular}
\caption{Many domain mesh refinement studies for Case 4}
\label{tab:meshRefinement3}
\end{table}

In addition to four cases mentioned, we also conducted the verification studies for generalized versions of case 2 and case 4 with 10 and 50 pellet domains to demonstrate the convergence rate of the solution procedure in parallel. These results are presented in Tables \ref{tab:meshRefinement2}-\ref{tab:meshRefinement3}.
\subsection{Validation Studies}
\label{sec:validation}

An extensive validation evaluation of AMP for nuclear fuel applications, which includes several experimental fuel rods for a variety of conditions, has been documented in \cite{phillippe_thesis, phillippe_2014a, phillippe_2014b}.  
This section includes an excerpt of that research to provide a basis for the accuracy of the material models with respect to the experimental results.  
Because of the extreme environment of nuclear fuel (high radiation, high temperature, and highly turbulent, multi-phase flow), it is difficult to precisely measure both the local temperature and the power in the fuel near the thermocouple.
Therefore, nuclear fuel experiments generally assume a 5 to 10\% experimental uncertainty in the measured data; for this report, we have incorporated a relatively tight expected tolerance of $\frac{+}{}$ 50K, which is generally less than the 10\% error and approximately 3\% at full power.

\begin{figure}
   \centering
   \includegraphics[scale=0.5]{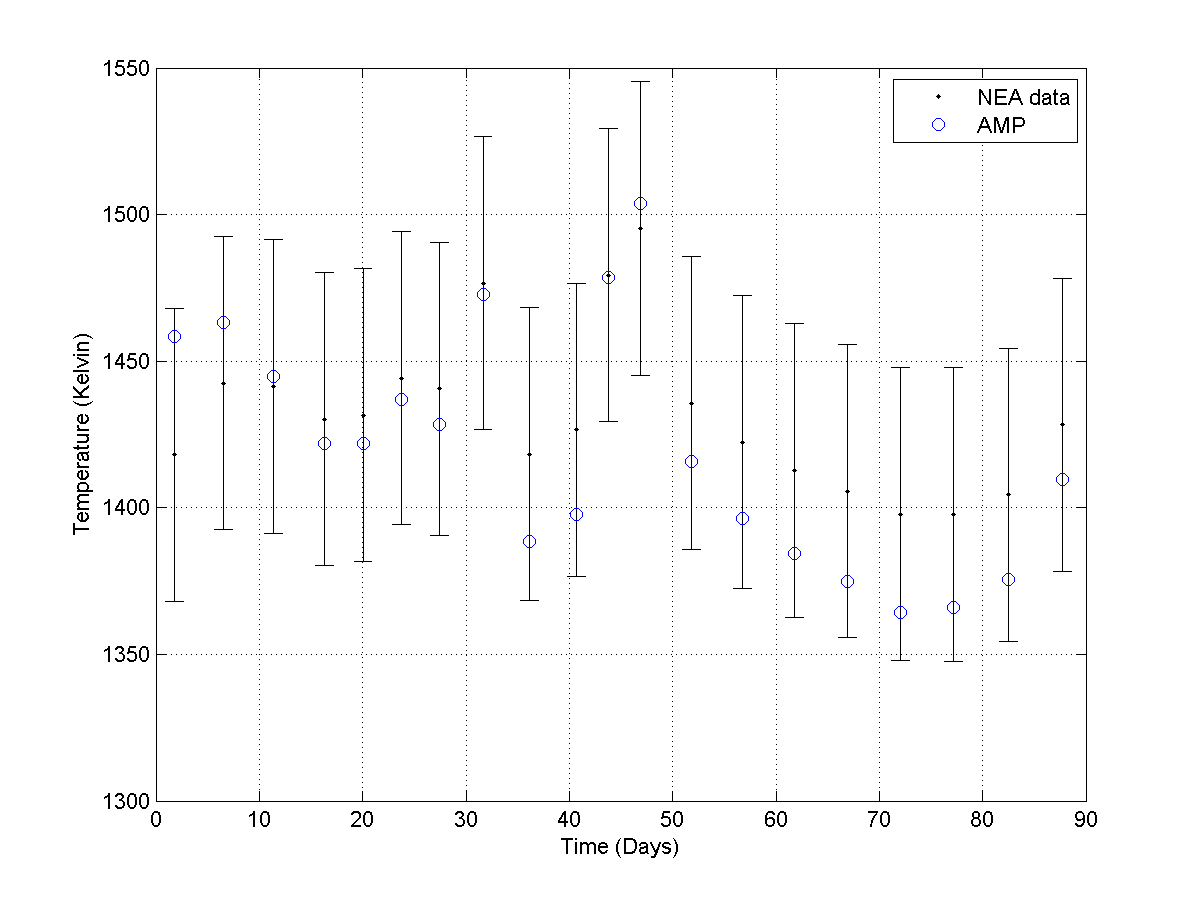}
   \caption{Validation of the temperature at the thermocouple in the IFA 597 experiment} 
   \label{fig:ifa597Results}
 \end{figure}

\begin{figure}
   \centering
   \includegraphics[scale=0.5]{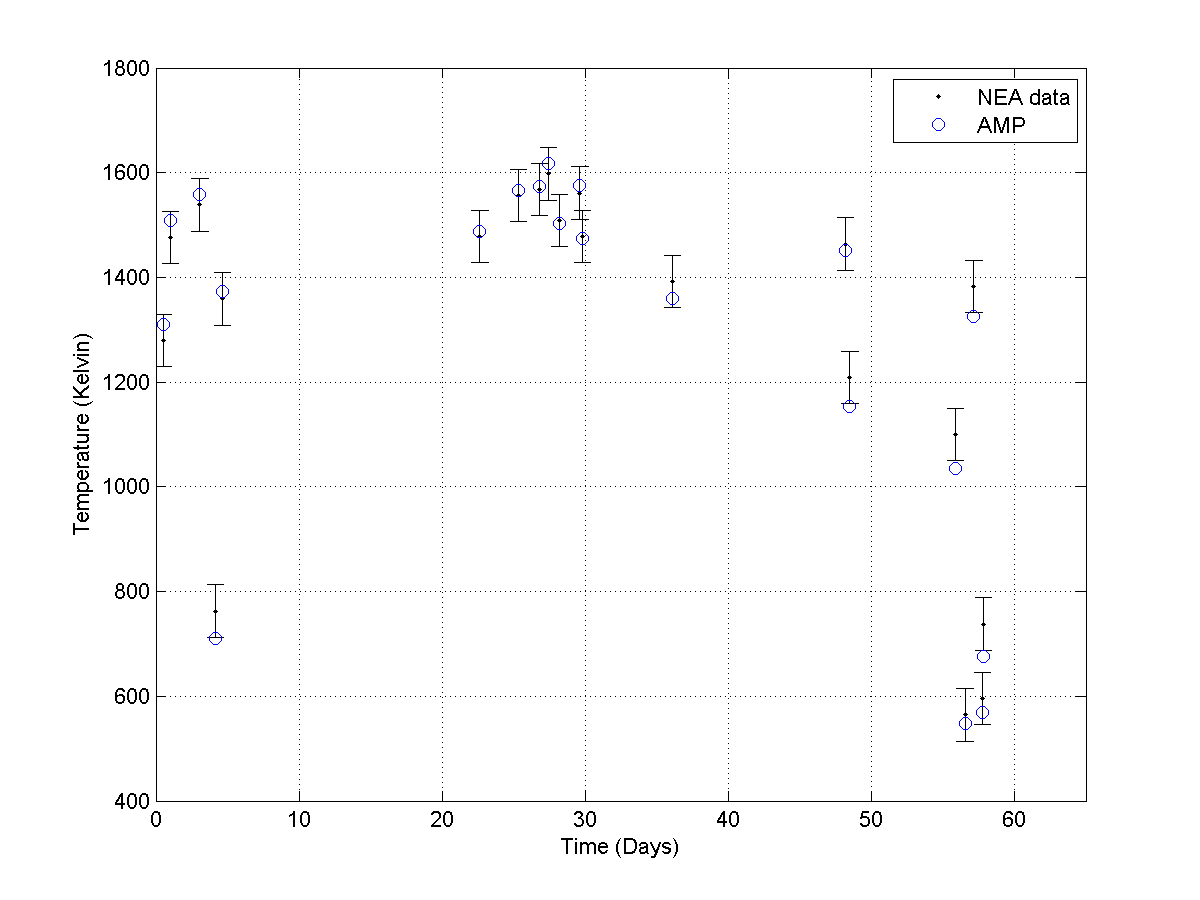}
   \caption{Validation of the temperature at the thermocouple in the IFA 432 experiment} 
   \label{fig:ifa432Results}
 \end{figure}

Figure \ref{fig:ifa597Results} provides the computational results of the AMP simulation and experimental measurements of the centerline fuel temperature in the IFA 597 experiment.  The input power is relatively constant and the computational results are consistently within 50K and generally  within 3\% of the measured temperature.  The results for the IFA 432 experiment are shown in Figure \ref{fig:ifa432Results}.  Because the power distribution varies significantly more than in the IFA 597 experiment, the 50K error bars on the experimental results appear much smaller.  However, the AMP results generally fall within the experimental error.  

\subsection{Scaling Studies}
\label{sec:scaling}

Scaling studies for the steady state thermal problem are presented here.  The problem setup was described in 
Section \ref{sec:testSetup}, with the geometry based on the IFA-432 experiment, but now including 348 full pellets, 
which is representative of a full-scale commercial nuclear fuel rod.  Several different aspects of the performance 
are studied using both strong and weak scaling.  These include:
\begin{itemize}
\item {\bf{Solve}}: Total time spent within the JFNK solver for the entire coupled nonlinear thermal problem. 
\item {\bf{Nonlinear Residual}}: The time required to compute one consistent global nonlinear residual within the JFNK solver across all physical domains. Computing the nonlinear residual involves computing the nonlinear residual for each physical volume, applying boundary conditions, and mapping solution components across domains. The times for these individual components will also be reported separately. Note that the number of calls will vary slightly depending on the number of iterations of the solver.  This information is also included.
\item {\bf{Diffusion Apply}}: Time required to compute the non-linear finite element residual at all interior degrees of freedom within all domains once. No communication between domains is required here.
\item {\bf{Boundary Conditions}}: Time required to impose the non-linear finite element boundary conditions once within the computation of the global nonlinear residual. This requires traversing all surface elements on the finite element meshes for each domain. No communication between domains is required here.
\item {\bf{Map Apply}}: Time per application of the maps for the solution transfer between the clad and pellets and the maps between the different pellet domains. The mapped values are required in imposition of the nonlinear Robin boundary conditions.
\item {\bf{Load Mesh}}: Time to create all of the meshes.  This consists of creating the appropriate load balance for the different domains, creating the individual meshes, and initializing all mesh data.  For the purposes of this problem the meshes used consist of internal, logically rectangular meshes.
\item {\bf{Save Results}}: Time to write the results.  This includes writing a separate SILO file for each core, and a single summary file that can be loaded into VisIt \cite{Childs:2005:ACS}.  
\end{itemize}

\subsubsection{Strong Scaling}
\label{sec:strongScaling}
The strong scaling results are based on a standard resolution mesh, which contains approximately 1.6M unknowns.  
For high fidelity analysis of a single fuel rod, this is considered a relatively coarse mesh; a high resolution 
analysis would typically require an 8x increase in each dimension.  On the other hand, for high-fidelity analyses of a full nuclear reactor 
containing tens of thousands of fuel rods, this resolution is sufficient because the accuracy requirements are typically lower.  
Demonstrating efficient strong scaling from serial to 100-500 cores for this problem size (single rod with $\sim$1.6M unknowns) will be sufficiently indicative 
of good scaling results for a full reactor on millions of cores and will also provide a lower bound on expected scaling for fine-resolution fuel 
rod calculations on tens of thousands of cores.  

\begin{figure}[!h]
\begin{center}
\begin{tikzpicture}
    \begin{loglogaxis}[
        xlabel=\# of Processors,
        ylabel=Time (s),
        xmin=1,
        xmax=2048,
        ymin=0.1,
        ymax=1e4,
        x=1.1cm,
        y=0.4cm,
        domain=1:2048,
        legend pos=outer north east,
        legend style={draw=none,font=\scriptsize}],
    ]
    \addplot [domain=1:2048] {9800/x};
    \addplot table[x index=0,y index=1] {strongscaling_titan.dat};
    \addplot table[x index=0,y index=2] {strongscaling_titan.dat};
    \addplot table[x index=0,y index=3] {strongscaling_titan.dat};
    \addplot table[x index=0,y index=4] {strongscaling_titan.dat};
    \addplot table[x index=0,y index=5] {strongscaling_titan.dat};
    \addplot table[x index=0,y index=6] {strongscaling_titan.dat};
    \addplot table[x index=0,y index=7] {strongscaling_titan.dat};
    \legend{ Ideal \\ Solve \\ Apply \\ Diffusion Apply \\ Robin Apply \\ Maps \\ Load \\ Save \\}
    \end{loglogaxis}
\end{tikzpicture}
\caption{Strong scaling studies on Titan}\label{fig:strongScalingFigureTitan}
\end{center}
\end{figure}
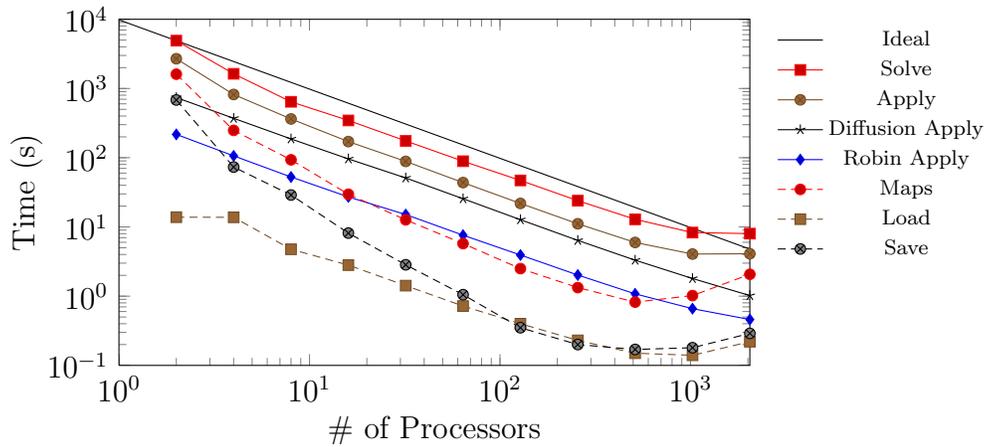

The strong scaling studies were executed using 1-2048 cores on the Titan Cray XK7 and the EOS XC30 at Oak Ridge National Laboratory.  
Tables \ref{table:strongScalingResultsEOS} and \ref{table:strongScalingResultsTitan} show the scaling results on EOS and Titan respectively, 
with the different components of the solve as well as loading the meshes and saving the results.  
The first column is the number of cores, while all other columns are the wall-clock execution times for the problem in seconds.  
For the diffusion and map apply calls (Diffusion Apply and Map Apply, respectively), the execution time is the average accumulated time across all cores.  
This is necessary because some cores may have different execution times due to load imbalances and is particularly acute for small core counts (2-8) 
due to a load imbalance between the domains.
For large core counts, each domain will exist on a non-overlapping set of cores.  
Figures \ref{fig:strongScalingFigureEOS} and \ref{fig:strongScalingFigureTitan} shows the plots of the scaling results compared to ideal scaling.


\begin{table}[!h]
\footnotesize
    \newcolumntype{C}{@{\extracolsep{-0.1cm}}>{\centering\arraybackslash}m{1.7cm}<{}}
    \pgfplotstabletypeset[
	    string type,
	    columns/procs/.style={column name=\# of Processors, column type={C}},
	    columns/solve/.style={column name=Solve, column type={C}},
	    columns/global/.style={column name=Nonlinear Residual, column  type={C}},
	    columns/diffusion/.style={column name=Diffusion Apply, column type={C}},
	    columns/robin/.style={column name=Boundary Conditions, column type={C}},
	    columns/maps/.style={column name=Map Apply, column type={C}},
	    columns/load/.style={column name=Mesh Loading, column type={C}},
	    columns/save/.style={column name=Save Results, column type={C}},
	    every head row/.style={before row=\hline,after row=\hline},
	    every last row/.style={after row=\hline},
	    ]{strongscaling_eos.dat}
    \caption{Strong scaling studies on EOS}
\label{table:strongScalingResultsEOS}
\end{table}

\begin{table}[!h]
\footnotesize
    \newcolumntype{C}{@{\extracolsep{-0.1cm}}>{\centering\arraybackslash}m{1.7cm}<{}}
    \pgfplotstabletypeset[
	    string type,
	    columns/procs/.style={column name=\# of Processors, column type={C}},
	    columns/solve/.style={column name=Solve, column type={C}},
	    columns/global/.style={column name=Nonlinear Residual, column  type={C}},
	    columns/diffusion/.style={column name=Diffusion Apply, column type={C}},
	    columns/robin/.style={column name=Boundary Conditions, column type={C}},
	    columns/maps/.style={column name=Map Apply, column type={C}},
	    columns/load/.style={column name=Mesh Loading, column type={C}},
	    columns/save/.style={column name=Save Results, column type={C}},
	    every head row/.style={before row=\hline,after row=\hline},
	    every last row/.style={after row=\hline},
	    ]{strongscaling_titan.dat}
    \caption{Strong scaling studies on Titan}
\label{table:strongScalingResultsTitan}
\end{table}

\begin{figure}[!h]
\begin{center}
\begin{tikzpicture}
    \begin{loglogaxis}[
        xlabel=\# of Processors,
        ylabel=Time (s),
        xmin=1,
        xmax=2048,
        ymin=0.1,
        ymax=1e4,
        x=1.1cm,
        y=0.4cm,
        domain=1:2048,
        legend pos=outer north east,
        legend style={draw=none,font=\scriptsize}],
    ]
    \addplot [domain=1:2048] {3488/x};
    \addplot table[x index=0,y index=1] {strongscaling_eos.dat};
    \addplot table[x index=0,y index=2] {strongscaling_eos.dat};
    \addplot table[x index=0,y index=3] {strongscaling_eos.dat};
    \addplot table[x index=0,y index=4] {strongscaling_eos.dat};
    \addplot table[x index=0,y index=5] {strongscaling_eos.dat};
    \addplot table[x index=0,y index=6] {strongscaling_eos.dat};
    \addplot table[x index=0,y index=7] {strongscaling_eos.dat};
    \legend{ Ideal \\ Solve \\ Apply \\ Diffusion Apply \\ Robin Apply \\ Maps \\ Load \\ Save \\}
    \end{loglogaxis}
\end{tikzpicture}
\caption{Strong scaling studies on EOS}\label{fig:strongScalingFigureEOS}
\end{center}
\end{figure}
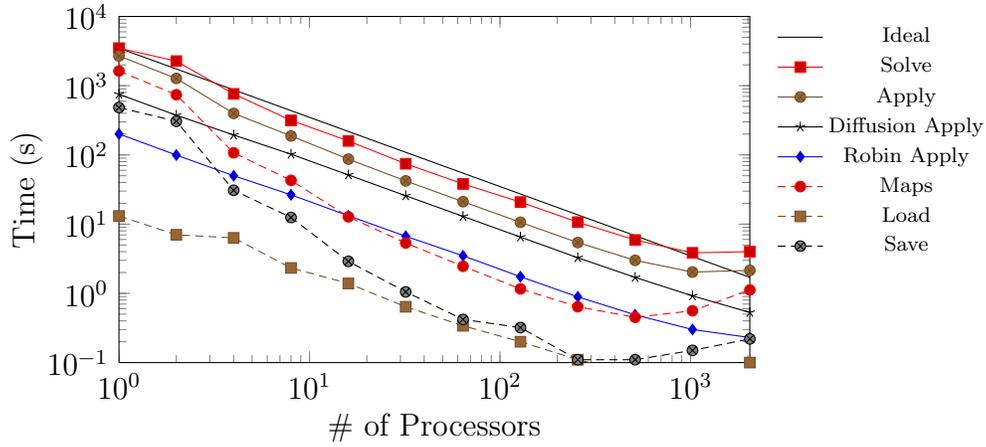

\afterpage{
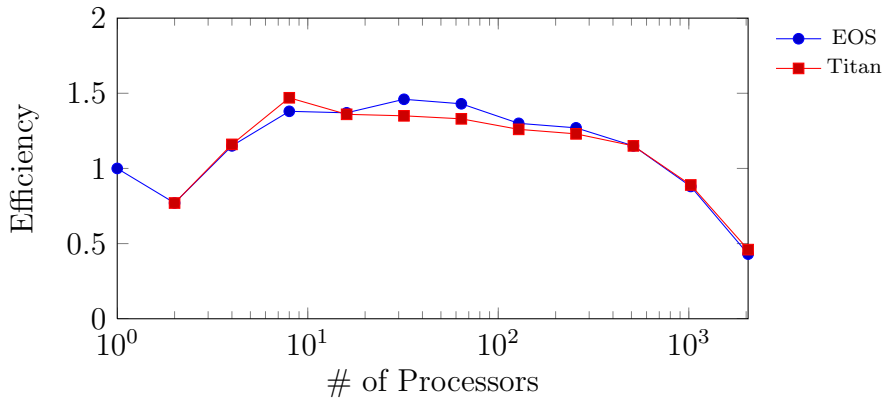
\begin{figure}[!h]
\begin{center}
\begin{tikzpicture}
    \begin{semilogxaxis}[
        xlabel=\# of Processors,
        ylabel=Efficiency,
        xmin=1,
        xmax=2048,
        ymin=0,
        ymax=2,
        x=1.1cm,
        y=2.0cm,
        domain=1:2048,
        legend pos=outer north east,
        legend style={draw=none,font=\scriptsize}],
    ]
    \addplot table[x index=0,y index=1] {scalingEfficiencyEos.dat};
    \addplot table[x index=0,y index=1] {scalingEfficiencyTitan.dat};
    \legend{ EOS \\ Titan \\}
    \end{semilogxaxis}
\end{tikzpicture}
\caption{Scaling efficiency\protect\footnotemark}\label{fig:scalingEfficiencyPlot}
\end{center}
\end{figure}
\footnotetext{The serial wall clock time $T_S$ for Titan is not available due to maximum wall clock time restrictions.  
$T_S$ on Titan is estimated by assuming that the efficiency on 2 processors
of Titan and EOS is similar.  This assumption is justified by the data in Tables
\ref{table:strongScalingResultsEOS} and \ref{table:strongScalingResultsTitan}. } 
}

For small core counts (1-8), the run time of the solve is limited by the maps between the different domains.  The most time-consuming map is for the clad-to-pellets heat transfer between the outer surface of the pellets and the inner surface of the clad.  This map is constructed using a std::multimap with all points on the local surface, followed by a pair-wise all-all exchange of data between the two meshes.  For the serial case, this results in a large number of points in the local map that must be managed.  As the number of cores is increased to 128-512 cores, the number of points on the surface per processor is significantly reduced, which leads to an additional log(n) reduction in the wall-clock time that results in the super-linear speedup observed.  The apparent lack of speedup from 1 to 2 cores is due to the load imbalance between the clad mesh located on one core and all of the pellet meshes on the other.  
For very large core counts (1024-2048), the number of elements on a surface is sufficiently small that the communication time begins to dominate, which limits the scaling for this problem size.  The behavior of the diffusion and the Robin boundary condition applies (Diffusion Apply and BC Apply, respectively) show nearly perfect scalability because they do not involve any communication.  
Generating the native structured meshes does not represent a significant portion of the execution time, yet loading the meshes shows nearly perfect scaling up to 512 cores.  At this relatively high core count, the load time is dominated by the load balance process, which is relatively independent of the number of cores ($\sim$0.1 seconds).  Finally, saving the results of the simulation (Save Results) has an acceptable scalability.  The results are saved to multiple SILO files (one per core), with a single summary file.  For small core counts, the Save Results is dominated by the time to write the data.  For large core counts, Save Results is dominated by the time to open a file on the Lustre file system.  This is approximately constant for all cores, but has a large variation that depends on the load of the computer.  Several executions were made and the typical results are presented.  The time to open a file varied between 0.02 and 1 second.  Large executions are particularly sensitive to this effect because all cores must synchronize to write the summary file. 
Figure \ref{fig:scalingEfficiencyPlot} shows the parallel efficiency on EOS and Titan.  
The efficiency is calculated as $\frac{T_S}{P_N*T_P}$, where $T_S$ is the serial wall clock time,
$P_N$ is the number of processors, and $T_P$ is the parallel wall clock time with $P_N$ processes.
Note that for processor counts between 4 and 512, the speedup is greater than 1.  
This is due to the super-linear speedup discussed previously.

\subsubsection{Weak Scaling}
\label{sec:weakScaling}
Tables \ref{table:weakScalingResultsEOS} and \ref{table:weakScalingResultsTitan} shows the results of a weak scaling study performed on EOS and Titan.  In the weak scaling study, the base mesh from the strong scaling study was used with 64 cores and the resolution was increased in each direction by a factor of 2x and 4x.  This resulted in a set of executions of $\sim$1.6M unknowns on 64 cores, $\sim$12.8M unknowns on 512 cores, and $\sim$102M unknowns on 4096 cores.  All times are in seconds and for functions that are called multiple times per solve the time per call is included in parentheses.

Using weak scaling, varying the problem size and processor count by 64 times, the performance of the solve is approximately constant.  The total number of non-linear iterations was independent of the resolution, and the total number of linear iterations varies slightly with resolution.  This is most likely due to a slight degradation in the performance of the parallel smoother used within the algebraic multigrid solver. The solve time is approximately constant with some variation that is due to differences in the number of linear iterations.  The contributions to the solve are specified in terms of the total time and the time per iteration in parentheses.  Based on the time per iteration, the global apply, finite element diffusion operator apply, and resetting of Trilinos ML are all approximately constant.  The apply call for the Robin boundary condition decreases slightly with problem size as the ratio of the total number of unknowns on the surface compared to the total number of unknowns decreases slowly with problem size.  The variation in mapping between domains is primarily due to specific parallel decomposition and the variation in MPI performance on Titan and EOS that depend on the allocated nodes.  Loading and saving the meshes show a slight increase for large problems due to increased demand on the Lustre parallel file system used but represents a small fraction of the total run time (2-3\%).  

\begin{table}[!h]
\centering
\begin{tabular}{|l|ccc|}
\hline
                        &       1x      &       2x      &      4x        \\
\hline
Core Count              &       64      &       512     &     4096       \\
Degrees of Freedom      &      1.6M     &      12.8M    &     102M       \\
Nonlinear iterations    &       5       &        5      &       5        \\
Linear iterations       &       29      &       31      &      36        \\
Solve                   &     38.14     &     35.29     &     49.56      \\
Nonlinear Residual      &  21.10 (0.73) &  19.24 (0.62) &  26.91 (0.75)  \\
Diffusion Apply         &  12.97 (0.45) &  14.04 (0.45) &  17.12 (0.48)  \\
Boundary Conditions     &   3.46 (0.12) &   2.00 (0.06) &   1.81 (0.05)  \\
Map Apply               &   2.46 (0.08) &   1.60 (0.05) &   5.97 (0.17)  \\
Reset ML                &  13.05 (0.45) &  11.22 (0.26) &  12.74 (0.35)  \\
Mesh Loading            &      0.34     &      0.28     &      0.42      \\
Save Results            &      0.42     &      0.35     &      1.11      \\
\hline
\end{tabular}
\caption{Weak scaling studies on EOS}
\label{table:weakScalingResultsEOS}
\end{table}

\begin{table}[!h]
\centering
\begin{tabular}{|l|ccc|}
\hline
                        &       1x      &       2x      &     4x        \\
\hline
Core Count              &       64      &       512     &     4096       \\
Degrees of Freedom      &      1.6M     &      12.8M    &     102M       \\
Nonlinear iterations    &       5       &        5      &       5        \\
Linear iterations       &       29      &       31      &      36        \\
Solve                   &     88.99     &     77.94     &    107.19      \\
Nonlinear Residual      &  43.66 (1.51) &  38.10 (1.23) &  55.69 (1.55)  \\
Diffusion Apply         &  25.55 (0.88) &  27.51 (0.89) &  37.12 (1.03)  \\
Boundary Conditions     &   7.57 (0.26) &   4.30 (0.14) &   4.09 (0.11)  \\
Map Apply               &   5.74 (0.20) &   3.05 (0.10) &  10.22 (0.28)  \\
Reset ML                &  29.27 (1.01) &  25.09 (0.81) &  30.54 (0.85)  \\
Mesh Loading            &      0.72     &      0.59     &      1.25      \\
Save Results            &      1.06     &      0.72     &      1.47      \\
\hline
\end{tabular}
\caption{Weak scaling studies on Titan}
\label{table:weakScalingResultsTitan}
\end{table}

\subsubsection{Preconditioner Performance}
\label{sec:domainScaling}
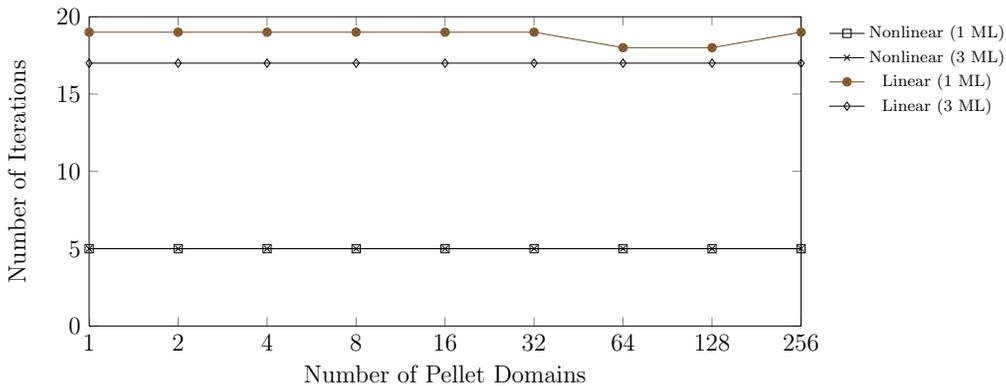
\begin{figure}[!h]
\centering
    \resizebox{\textwidth}{!}{
\begin{tikzpicture}
    \begin{axis}[
        xlabel=Number of Pellet Domains,
        ylabel=Number of Iterations,
        xmin=1,
        xmax=256,
        xmode=log,
        log basis x={2},
        xticklabels={1,2,4,8,16,32,64,128,256},
        ymin=0,
        ymax=20,
        height=0.5\textwidth,
        width=\textwidth,
        legend pos=outer north east,
        legend style={draw=none,font=\scriptsize}],
    ]
    \addplot[mark=square] table[x index=0,y index=1] {domainIterations.dat};
    \addplot[mark=x] table[x index=0,y index=2] {domainIterations.dat};
    \addplot table[x index=0,y index=3] {domainIterations.dat};
    \addplot[mark=diamond] table[x index=0,y index=4] {domainIterations.dat};
    \legend{ Nonlinear (1 ML) \\ Nonlinear (3 ML) \\ Linear (1 ML) \\ Linear (3 ML) \\}
    \end{axis}
\end{tikzpicture} 
}
\caption{Number of nonlinear and linear iterations as a function of the number of pellet domains and the number of algebraic multigrid (ML) iterations}\label{fig:domainScalingResults}
\end{figure}

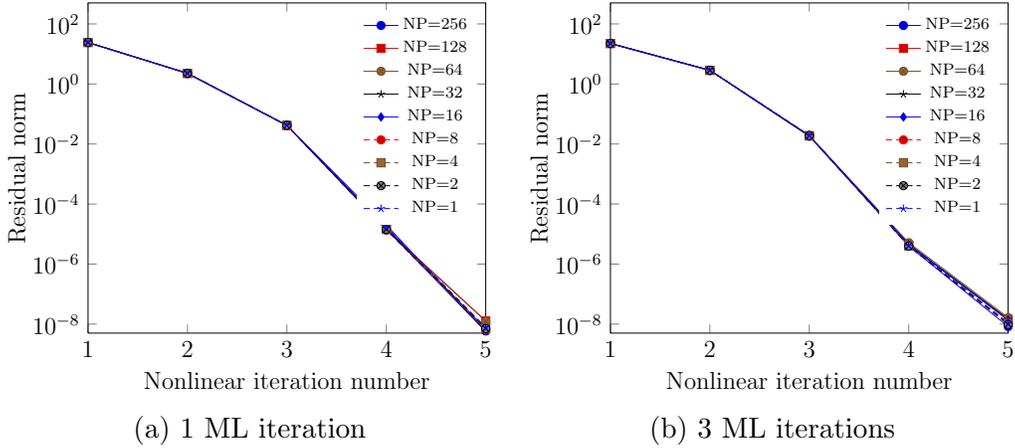
\begin{figure}[ht]
    \begin{subfigure}[b]{0.5\textwidth}
    \centering
    \resizebox{\textwidth}{!}{
    \begin{tikzpicture}
    \begin{semilogyaxis}[
        xlabel=Nonlinear iteration number,
        ylabel=Residual norm,
        xmin=1,
        xmax=5,
        ymin=5e-9,
        ymax=5e2,
        legend style={draw=none,font=\scriptsize},
    ]
    \addplot table[x index=0,y index=9] {domainConvergence1ML.dat}; \addlegendentry{NP=256}
    \addplot table[x index=0,y index=8] {domainConvergence1ML.dat}; \addlegendentry{NP=128}
    \addplot table[x index=0,y index=7] {domainConvergence1ML.dat}; \addlegendentry{NP=64}
    \addplot table[x index=0,y index=6] {domainConvergence1ML.dat}; \addlegendentry{NP=32}
    \addplot table[x index=0,y index=5] {domainConvergence1ML.dat}; \addlegendentry{NP=16}
    \addplot table[x index=0,y index=4] {domainConvergence1ML.dat}; \addlegendentry{NP=8}
    \addplot table[x index=0,y index=3] {domainConvergence1ML.dat}; \addlegendentry{NP=4}
    \addplot table[x index=0,y index=2] {domainConvergence1ML.dat}; \addlegendentry{NP=2}
    \addplot table[x index=0,y index=1] {domainConvergence1ML.dat}; \addlegendentry{NP=1}
    \end{semilogyaxis}
    \end{tikzpicture}
    }
    \caption{1 ML iteration}
    \end{subfigure}
    \hfill
    \begin{subfigure}[b]{0.5\textwidth}
    \resizebox{\textwidth}{!}{
    \begin{tikzpicture}
    \begin{semilogyaxis}[
        xlabel=Nonlinear iteration number,
        ylabel=Residual norm,
        xmin=1,
        xmax=5,
        ymin=5e-9,
        ymax=5e2,
        legend style={draw=none,font=\scriptsize},
    ]
    \addplot table[x index=0,y index=9] {domainConvergence3ML.dat}; \addlegendentry{NP=256}
    \addplot table[x index=0,y index=8] {domainConvergence3ML.dat}; \addlegendentry{NP=128}
    \addplot table[x index=0,y index=7] {domainConvergence3ML.dat}; \addlegendentry{NP=64}
    \addplot table[x index=0,y index=6] {domainConvergence3ML.dat}; \addlegendentry{NP=32}
    \addplot table[x index=0,y index=5] {domainConvergence3ML.dat}; \addlegendentry{NP=16}
    \addplot table[x index=0,y index=4] {domainConvergence3ML.dat}; \addlegendentry{NP=8}
    \addplot table[x index=0,y index=3] {domainConvergence3ML.dat}; \addlegendentry{NP=4}
    \addplot table[x index=0,y index=2] {domainConvergence3ML.dat}; \addlegendentry{NP=2}
    \addplot table[x index=0,y index=1] {domainConvergence3ML.dat}; \addlegendentry{NP=1}
    \end{semilogyaxis}
    \end{tikzpicture}
    }
    \caption{3 ML iterations}
    \end{subfigure}
    \caption{Residual convergence history for fuel rods with NP=1 to NP=256 pellet subdomains with 1 and 3 multigrid iterations per subdomain within the preconditioner}\label{fig:ConvergenceHistory}%
\end{figure}

\begin{table}
\small
\begin{tabular}{|c|c|c|c|c|c|c|c|c|c|}
\hline
Domains & 1 & 2 & 4 & 8 & 16 & 32 & 64 & 128 & 256 \\
\hline
Solve & 50.61 & 50.86 & 50.61 & 50.51 & 50.81 & 49.63 & 49.56 & 50.26 & 51.38 \\
\hline
\end{tabular}
\caption{Solution times as the number of pellet domains is varied}
\label{solveTimes}
\end{table}
As mentioned in the introduction section the number of pellets within a fuel rod can vary dramatically. 
Hence it is important that the solver deliver good performance as the number of actual pellet domains is varied. 
This in turn is dependent primarily on the performance of the preconditioner employed. In order to study this 
we consider a series of numerical experiments where the total number of mesh elements across all pellet and 
clad domains is kept constant while the number of pellet domains is varied. We note that though the number of 
mesh elements is kept constant the number of degrees of freedom does rise slightly (3\%) as the number of pellets 
is increased due to the introduction of more surface elements. In addition the number of solution transfer operations that 
require communication between domains also increases. Note that the additional communication incurred is point to
point communication between pairs of processors and does not significantly affect the runtime.
The dimensions of the pellets are chosen so that the height of the pellet stack matches that of the clad. 
Figure \ref{fig:domainScalingResults} plots the number of nonlinear and linear iterations required for solving 
thermal fuel rod problems as the number of pellet domains is varied from 1 through 256. In addition the number 
of multigrid solves required within the preconditioner is varied from 1 to 3. As can be seen, the number of 
iterations remains largely constant. Figures \ref{fig:ConvergenceHistory}(a) and  \ref{fig:ConvergenceHistory}(b) 
show the residual convergence history for the nonlinear solver as the number of domains is varied ranging from 1-256 domains. 
There is only a very slight effect on the residual due to increasing the number of domains. Finally, Table \ref{solveTimes} 
shows that the required wall clock times for solution as the number of domains is varied does not change significantly.

\section{Fuel Assembly Modeling}
\label{sec:assembly}
In section \ref{sec:infrastructure} the components of the AMP multi-physics infrastructure that enabled the development of nonlinearly consistent multi-domain thermal transport calculations that form the main core of this paper were described. Here we illustrate further multi-physics capabilities of AMP by describing further extensions of the fuel rod modeling capability. Since our focus is on solution and coupling methodology and due to space limitations we will concentrate on the relevant coupling aspects with details on the models being provided in the appendices and provided references.\\
\noindent {\it Coupling to Coolant Models: } The coolant liquid flowing on the outside of each fuel rod serves as a heat sink which flows axially along the length of
the outside surface of the clad. The coolant model in the fluid domain and the thermal transport model in the clad domain are coupled nonlinearly through
Robin boundary conditions (Eqn \ref{C2FRobinBC})
\begin{equation*}
k_c(T_c)\nabla T_{c}\cdot \mathbf{n_f} + h_{c,f}(T_c, T^m_f) (T_{c} - T^m_{f} )  = 0 \; \text{on} \;\Gamma_{c,f} \\
\end{equation*}
reproduced here for clarity. Here, $T^m_f$ represents a mapped coolant temperature field over the clad surface.
AMP enables us to couple different models interchangeably.
The first model described in \ref{sec:single_eq_flow} uses a reduced empirical model that solves a single axial equation using a simple finite difference scheme and
is frequently sufficient for many calculations. This is the model used within the single fuel rod calculations presented in prior sections. The second model described in
\ref{sec:subchannel_flow}  solves the two equations using a more complex model that is used when subchannel temperatures and densities are required. This model
is used in the fuel assembly calculations that we will describe further along. Let
\begin{equation}
F_f(T^h_f) = 0
\label{NLFlowModel}
\end{equation}
denote the nonlinear system of equations resulting from discretizing either coolant model over the fluid domain with $T^h_f$ a vector of fluid temperature unknowns. As described in Section \ref{sec:jfnk}, a JFNK solver only requires us to provide the ability to compute a nonlinear residual. Hence, augmenting the existing nonlinear system (\refeq{eqn::FullNLE}) of equations across the pellet and clad domains with eqns (\ref{NLFlowModel}) for the flow domain enables us to perform coupled flow and thermal transport calculations. The augmented nonlinear system can be denoted by:
\begin{equation}
\mathbf{\tilde{F}}(\mathbf{T})  = \mathbf{0}
\label{eqn::FullFlowNLE}
\end{equation}
 where $\mathbf{\tilde{F}}(\mathbf{T}) = \left( F_1(T_1^h), F_2(T_2^h), \ldots, F_N(T^h_N), F_c(T^h_c), F_f(T^h_f) \right)^t$ is the coupled set of block nonlinear equations across all pellet, clad, and flow domains with $\mathbf{T}$ denotes  the block unknowns across all domains including flow. For preconditioning an augmented approximate Jacobian system of the form
 \begin{equation}
 \mathbf{\tilde{M} } = \begin{bmatrix}
\mathbf{M} & 0 \\
0         & \tilde{J}_{ff} \\
\end{bmatrix}
 \end{equation}
is used with $\mathbf{M}$ as defined in Eqn (\ref{PCSystem}) and $\tilde{J}_{ff}$ an approximate Jacobian for the flow model. Since reduced order models are used for the flow, systems involving $\tilde{J}_{ff}$ are inverted using a direct solver as opposed to the multigrid solvers used to invert the other components. We note that the direct solver for the flow operates in parallel to the multigrid solves during each preconditioner solve step.\\

\noindent {\it Coupling to Oxide Growth Models: }
The formation of an oxide layer on the surface of a nuclear fuel rod can interfere with
thermal conduction to the coolant and create additional stress within the clad. Typically the thickness of the oxide layer is much smaller
than the other physical scales in the fuel rod. As a result the oxide growth is modeled
at each point, $\mathbf{x}$, on the surface of the clad by independent 1D models of the form:
\begin{equation}
    \displaystyle \frac{\partial C(\mathbf{x},t)}{\partial t} =
    \frac{\partial}{\partial x}\left[D(\mathbf{x})\frac{\partial C(\mathbf{x},t)}{\partial x}\right]
\label{1DOxideModel}
\end{equation}
Solving for the oxygen concentration then reduces to solving equation \ref{1DOxideModel} at each point on the surface of the clad
subject to appropriate initial and boundary conditions which is described in detail in \ref{sec:OxideModeling}.

For the purposes of this paper we consider a one way coupling of the full thermal transport model
to the oxide models. As described previously a nonlinear solve is performed to convergence over the pellets, clad, and coolant domains. The resulting temperature field is used to initialize the oxide growth models at each point on an exterior clad surface mesh generated from the clad mesh using the mesh subset capability of AMP. Each point on the mesh contains an independent sub-grid oxide model
which is distributed across the processors to match the clad load balance.  For the results
presented the clad runs on approximately half of the processors, with roughly the same number of surface
nodes on each processor.  Additionally since we have an independent model for every point
the problem is embarrassingly parallel between the points and no additional communication
is required.

\begin{figure}[ht]
\centering
\includegraphics[scale=0.35]{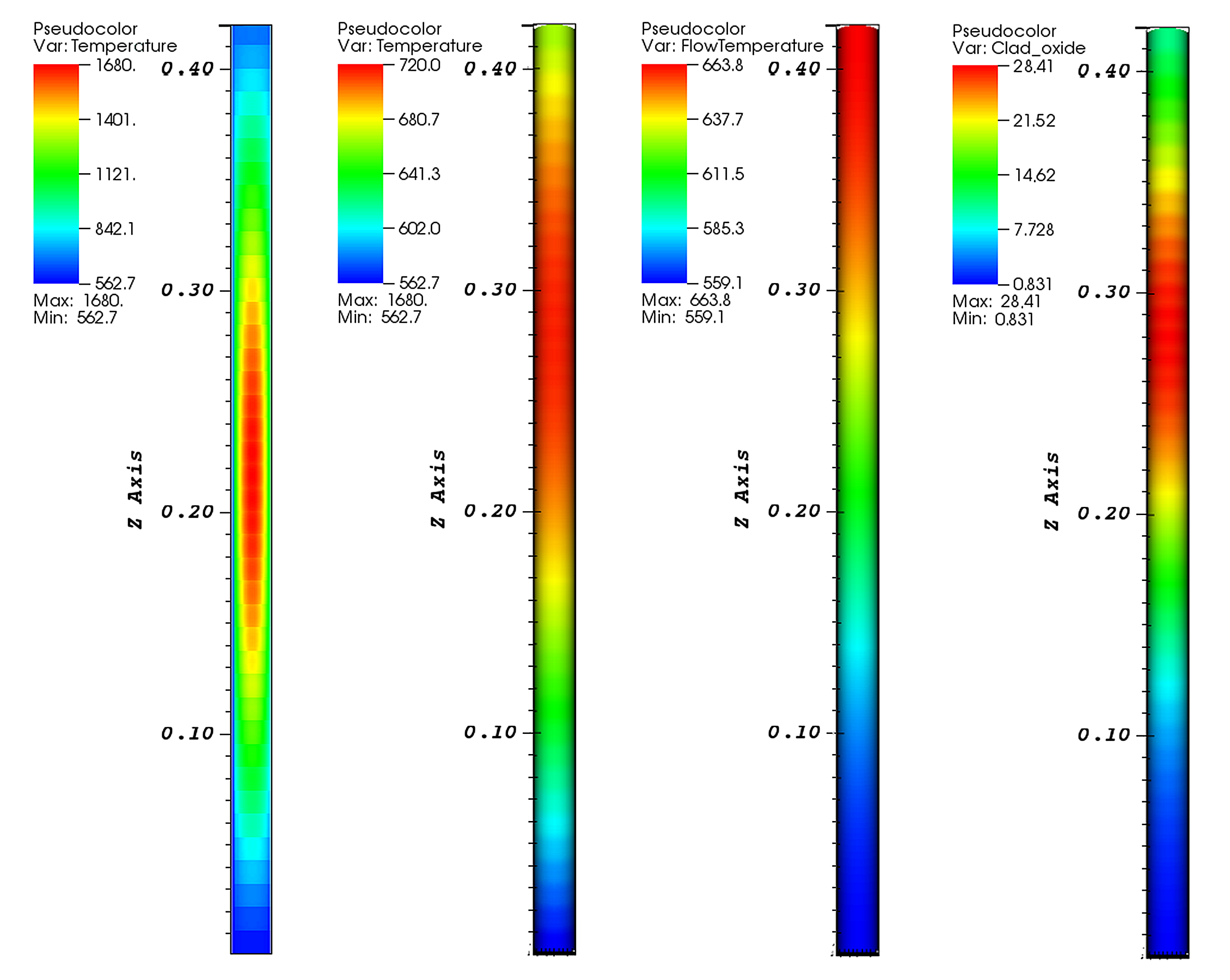}
\caption{ Oxide model results from left to right:
    Nuclear fuel pellet temperatures;
    Clad outer surface temperature;
    Coolant temperature;
    Resulting oxide layer thickness.
}
\label{fig:oxide}
\end{figure}

The results of the oxide model coupled to the thermal model is shown in Figure \ref{fig:oxide}.
The oxide layer thickness follows the surface temperature of the clad.  This temperature is
in turn affected by the power shape which can be seen by the fuel temperature and the flow temperature
which creates a top-shifted peak to the temperature of the clad and oxide layer thickness.  \\

\noindent {\it Coupled Radiation Transport and Fuel Assembly Thermal Modeling:} \\

In this section we describe the extension of the fuel rod modeling capability described to modeling
a full nuclear assembly and coupling with a massively parallel radiation transport code, Denovo  \cite{evans_2010}. Here we focus on describing
the coupling and scaling aspects and the interested reader is referred to
\cite{clarno_2012,hamilton_2012,hamilton_2013,phillippe_2014a,phillippe_2014b} for further details including detailed verification and validation
studies.

Solving the fuel assembly problem consists of two parts: solving the neutronics equations
to obtain the spatially varying source and solving the thermal diffusion with
the appropriate heat sink.  To accomplish this we solve the assembly level
radiation transport equations using the methods described in section \ref{sec:rad_trans_model},
and an array of multiple fuel rod problems each of which is solved as described in this
document.  To accomplish the latter, we utilize a multi-mesh capability that can replicate
a mesh to produce the appropriate array, and replicated column operators to produce
a complete nonlinear system.  The full nonlinear system is then solved using JFNK, with
the preconditioner limited to the individual sub domains.  The individual fuel rod physics are
embarrassingly parallel between the fuel rods which we utilize through our load balance by generating
a set of independent communicators for each fuel rod, with the fuel rods distributed on independent sets
of processors. We are not limited to this choice of parallel decomposition, but it yields
the best performance for this problem.

To demonstrate the ability to solve nuclear reactor analysis problems, we model a single
fuel assembly of a pressurized water reactor with coupling between heat transfer, subchannel
flow, and radiation transport.  This corresponds to CASL AMA Progression Problem 6 \cite{CASL_Prob6}.
The assembly consists of a $17 \times 17$ array of fuel rods on a square 1.26 cm pitch.
Of the 289 pins, 264 are fuel rods containing 3.1\%
enriched UO$_2$, 24 are guide tube locations, and a central instrumentation tube.  Zircaloy 4 clad
surrounds all pins.  The coolant surrounding the pins is water containing 1300 ppm soluble boron
and an inlet temperature of 569 K.  The average power level in the assembly is 30,000 W/kg,
approximately corresponding to an average power assembly from a full reactor core.
Further details on the geometry and material specifications can be found in \cite{CASL_Prob6}.

In the AMP computational model, the mesh for each fuel pellet contains 512 mesh
cells each fuel rod contains 360 pellets.  Additionally, each fuel rod is
surrounded by a clad mesh containing 54,144 cells for a total of 238,464 cells
per fuel rod.  Over the 264 fuel pins in the full assembly, the total number of mesh
cells in the AMP problem is approximately 63 million and the number of nodal
degrees of freedom is slightly over 100 million.
The Denovo computational model has approximately 4.6 million spatial cells, 23
energy groups, 32 angles, and P$_1$ scattering (which uses four angular moments).
The 23 group cross sections are collapsed from a 56 group library by the XSProc
module of the SCALE package \cite{SCALE}.
Power distributions computed by Denovo are mapped onto the AMP mesh using the polynomial smoothing process described
in \ref{sec:rad_trans_model}.  Temperatures and fluid densities computed by AMP
are averaged over each of 49 axial levels within every fuel rod to be used for generating new
cross sections.  A simple Picard iteration is used to couple AMP and Denovo, with a damping
(under-relaxation) factor of 0.4 applied to the temperature distribution, as described in
Ref.~\cite{hamilton_2013}.  The AMP thermal and subchannel problems are solved together using
a JFNK approach as described in \ref{sec:Coolant} and the Denovo $k$-eigenvalue problem is solved
using a Krylov-Schur eigensolver \cite{stewart_2001}.  A stopping criteria of $10^{-4}$ is
applied to the Picard iterations and a tolerance of $10^{-5}$ for each of the AMP and
Denovo subproblems.  The problem was decomposed across 4624 computational cores, with
both the AMP and Denovo problems utilizing the entire set of processors.  The coupled problem
converged in 12 Picard iterations, with an average of 1.6 Newton iterations per Picard iteration and 20 linear iterations per Newton step.  The entire solution required
3976 seconds, of which 3182 seconds were spent in the Denovo transport solves
and 680 seconds were spent in the AMP thermal solves.

\begin{figure}[htp]
\centering
    \begin{subfigure}[b]{0.38\textwidth}
    \includegraphics[width=\textwidth]{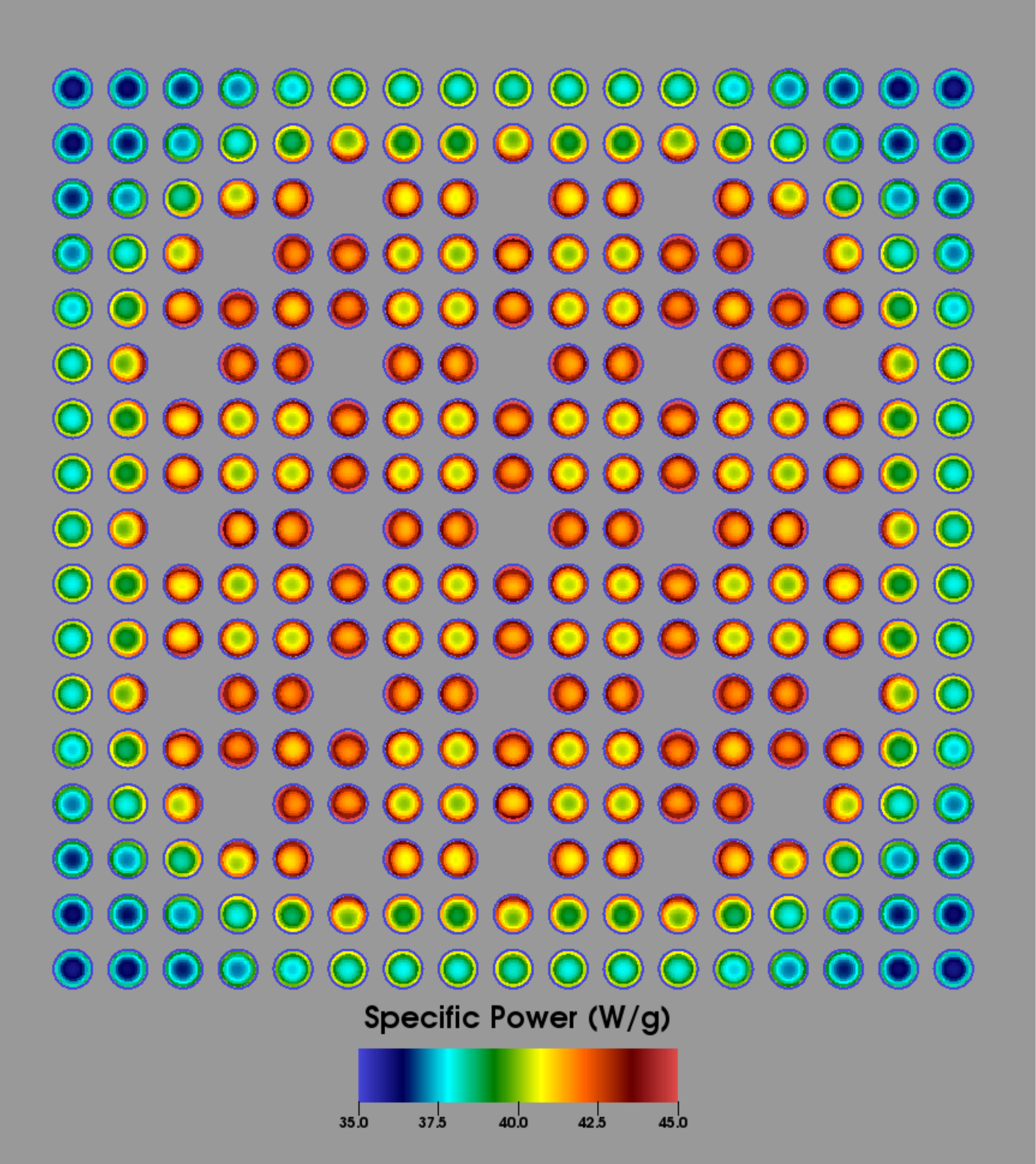}
	\caption{Radial Power Profile}
    \end{subfigure}
    \hfill
    \begin{subfigure}[b]{0.38\textwidth}
    \includegraphics[width=\textwidth]{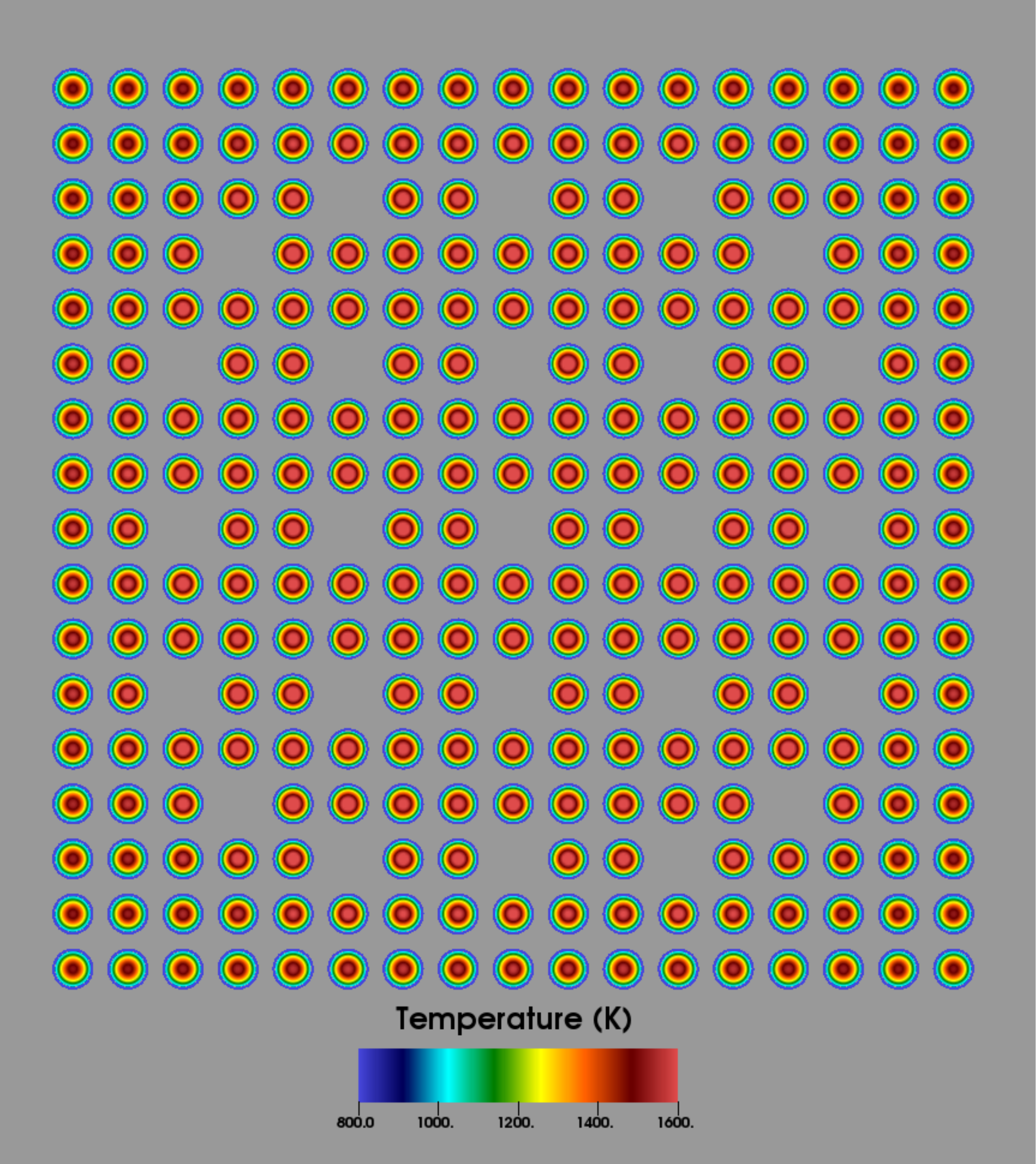}
    \caption{Radial Temperature Profile}
    \end{subfigure}
    \begin{subfigure}[b]{0.38\textwidth}
     \includegraphics[width=\textwidth]{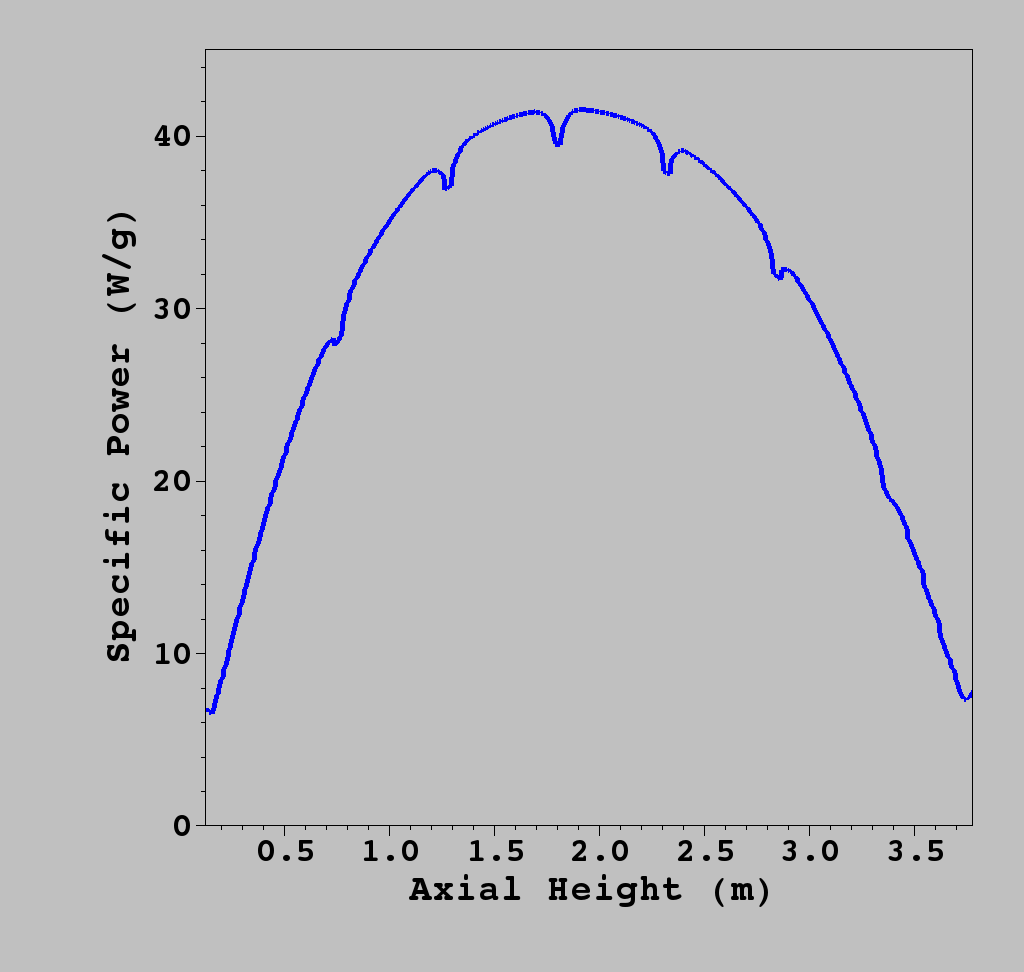}
    \caption{Axial Power Profile}
    \end{subfigure}
    \hfill
    \begin{subfigure}[b]{0.38\textwidth}
    \includegraphics[width=\textwidth]{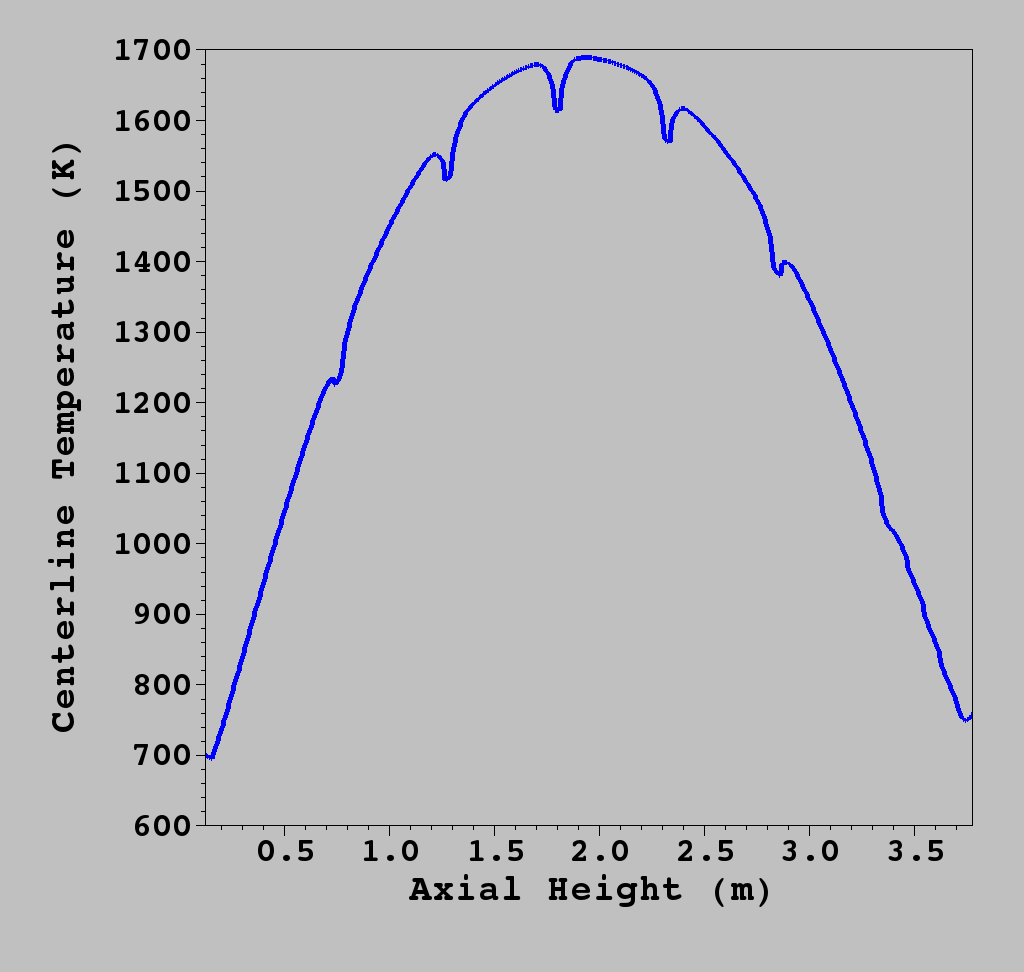}
    \caption{Axial Fuel Temperature Profile}
    \end{subfigure}
    \begin{subfigure}[b]{0.38\textwidth}
    \includegraphics[width=\textwidth]{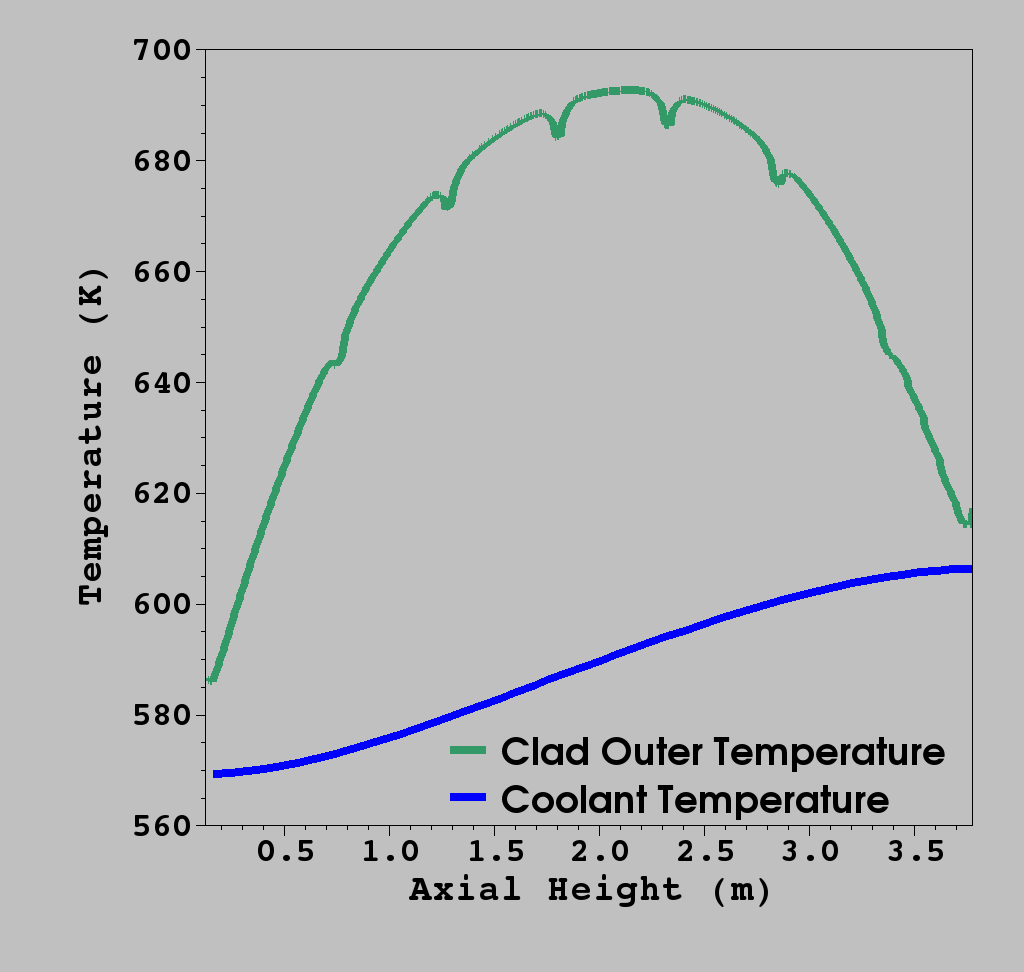}
    \caption{Clad and Coolant Axial Profiles}
    \end{subfigure}
    \hfill
    \begin{subfigure}[b]{0.38\textwidth}
    \includegraphics[width=\textwidth]{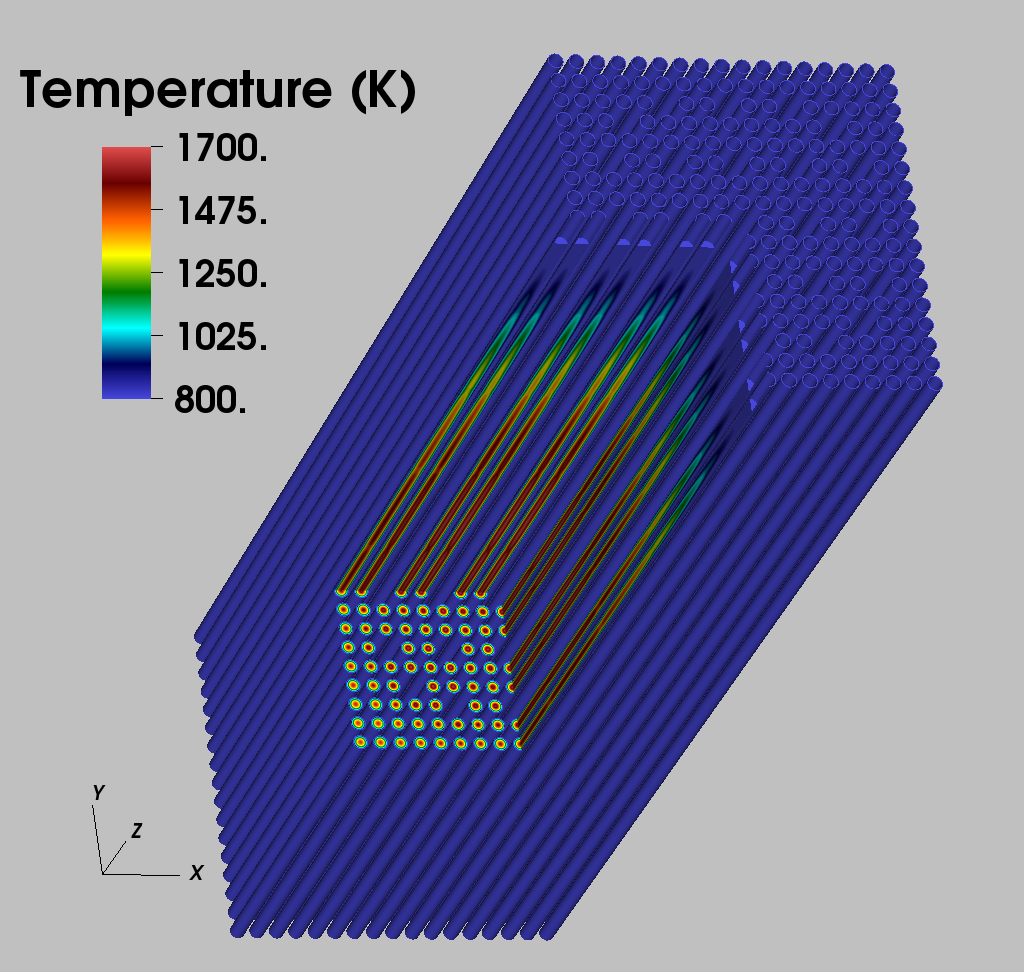}
    \caption{3D Temperature Profile}
    \end{subfigure}
\caption{Assembly Solution Details \label{fig:assby_viz}}
\end{figure}

The temperature and power solution profiles throughout the assembly are shown in
Fig.~\ref{fig:assby_viz}.
The radial variation of the power distribution, both within
individual fuel rods and across the assembly, is evident.  Notably, the power level in pins
neighboring guide tube locations is significantly higher than regions not near guide tubes
(such as the assembly corners) due to increased neutron moderation.  Although more difficult
to visually discern, the fuel temperature distributions mirrors the same general trends as
the power, with higher temperatures corresponding to high power regions.  Unlike the power,
which attains its peak values at the outer radius of the fuel rods, the temperature distribution
always peaks at or near the center of a fuel rod.
The axial profiles clearly show the presence of the spacer grids as a series of local
depressions due to increased absorption in those regions.

\section{Conclusions}
\label{sec:conclusion}
Many real world engineering problems involve the complex interaction between many bodies, in a nonlinear manner.  
From mesh generation to predicting results, modeling these large complex systems presents significant computational challenges.
An efficient parallel, multi-domain solution methodology has been developed and implemented to solve these systems by leveraging the natural decomposition of the problem associated with the individual domains.
This methodology has been demonstrated for modeling heat transfer within nuclear fuel rods, which are composed of hundreds of individual pellets within a metal tube, and can be applied to a nuclear reactor, which is composed of tens of thousands of individual fuel pins.

The model and discretization for the thermal transport in a nuclear fuel rod demonstration problem consists of nonlinear diffusion in each of the fuel pellets and clad, along with a Robin boundary condition on each surface, and maps between them, to account for the heat transfer between domains.
Modeling the entire system in a single domain would create a significant challenge associated with mesh generation, but by modeling many individual domains, the generation of the mesh for the full problem becomes automatic and a negligible burden on the user.

The Jacobian-Free Newton-Krylov (JFNK) method used to solve the nonlinear system of algebraic equations was described in detail, with a particular focus on the preconditioning strategy for the multi-domain problem.  
The Krylov solver has been shown to efficiently solve for the interaction between the individual domains, but an efficient preconditioner was required to account for the diffusion within individual domains.
Therefore, the preconditioning algorithm leverages the natural decomposition in physical space through the use of a parallel block-diagonal structure in which each block is a single, physical domain that neglects the interaction between domains.  

To efficiently solve these multi-domain problems, the computational infrastructure of AMP was designed to support the problem specification, domain decomposition, and composition of mathematical constructs efficiently in parallel.  
The computational infrastructure was described in detail, with a particular focus on the parallel implementation.
The infrastructure is designed to allow for the specification of nonlinear operators and linear preconditioners for individual domains and the composition of those operators into a single column operator that can be used by a solver, either internal or external, to compute the solution of the nonlinear problem.
Similarly, the parallel vectors (solution and source) can be defined for individual domains and composed into a single vector that is used by the solvers and preconditioners.
A ``subsetting'' approach has been developed to access individual components of the parallel vector that are associated with a single mesh (or surface of a mesh) and variable (such as temperature).
Because there is a knowledge of the individual physical domains, parallel communicators are created for each domain and the interacting surfaces between domains.
Therefore, the infrastructure can easily ``subset'' the parallel vector associated with an individual domain, determine a communicator specific to that domain, and access the preconditioned for that domain, which is provided to a solver for the compact preconditioning step on an individual domain.
The domain decomposition strategy is designed with an awareness of the individual domains to minimize communication during the computationally-expensive preconditioning and apply processes.

A series of numerical experiments were developed to verify, validate, and evaluate the parallel performance of the software and algorithm.
Each of the numerical experiments were based on a single specification of material properties and geometry that was associated with a nuclear fuel performance experiment that is used as an international benchmark for fuel performance modeling.
The numerical verification studies used the method of manufactured solutions and mesh refinement studies to verify the solution converged with the proper order to the manufactured solution.
The validation study, specified in much greater detail in an associated manuscript, demonstrated that the full problem, from material model specification to discretization and solution, could accurately predict the temperature distribution within a nuclear fuel rod within the experimental uncertainty bounds.  
Both strong and weak scaling studies were performed with significant profiling, to understand the performance of the software, and algorithm.  
Excellent strong scaling was achieved as the processor count increased three orders of magnitude, until the number of elements per core was under 1000.

This solution strategy, and the associated software, has been shown to accurately and efficiently solve large, complex, multi-domain, nonlinear problems by leveraging the natural structure of the physical domains.  
This has the potential to impact many computational engineering applications beyond nuclear fuel simulation.
\section*{Acknowledgements and Access}
The AMP (Advanced Multi-Physics) code is distributed with a modified BSD license and accessible either
by contacting the corresponding author or through the Radiation Safety Information Computational
Center (RSICC) at Oak Ridge National Laboratory, with an RSICC license, as CCC-793.
The development of AMP, and the nuclear fuel performance application built upon it, was funded by the Nuclear
 Energy Advanced Modeling and Simulation (NEAMS) program of the U.S. Department of Energy Office of Nuclear
 Energy, Advanced Modeling and Simulation Office. This material is also based upon work supported by the U.S. Department of Energy, Office of Science, Office of Advanced Scientific Computing Research, Applied Mathematics program, Extreme Scale Solvers project (EASIR).

Much appreciation is due to Aaron Phillippe, Jim Banfield, and Larry Ott for validation studies,
to William Cochran, Srdjan Simunovic, Jay Billings, and Phani Nukala for their early contributions to the software,
and to Mark Baird for computational support with the third party libraries.

This research used resources of the Oak Ridge Leadership Computing Facility at the Oak Ridge National Laboratory, which
 is supported by the Office of Science of the U.S. Department of Energy under Contract No. DE-AC05-00OR22725.
Mark Berrill acknowledges support from the Eugene P. Wigner Fellowship at Oak Ridge National Laboratory, managed
 by UT-Battelle, LLC, for the U.S. Department of Energy under Contract DE-AC05-00OR22725. 
\bibliographystyle{elsarticle/elsarticle-num}
\bibliography{refs}

\appendix

\section{Coolant Modeling:} 
\label{sec:Coolant}

The coolant liquid serves as a heat sink which flows axially along the length of 
the outside of the clad. Therefore simulating the coolant flow serves as a 
boundary condition for the thermal solve on the clad surface.  
Solving the coolant flow involves solving the fluid equations for conservation of mass,
momentum, and energy equations:
\begin{align}
& \frac{\partial \rho}{\partial t} + \nabla \cdot \left(\rho \vec{v} \right) = 0  \\
& \frac{\partial \rho v_i}{\partial t} = -\nabla \cdot \left(\rho v_i \vec{v} \right) +
    \left(-\nabla p + \nabla \cdot \vec{\tau} \right) - \vec{g}  \\
& \frac{\partial U}{\partial t} + \nabla \cdot \left(U \vec{v} \right) =
    -p \nabla \cdot \vec{v} + \Phi + \nabla k(T) \nabla T + \dot{q}
\end{align}
where $\rho$ is the mass density, $\vec{v}$ is the velocity, p is the pressure, $\vec{g}$ is
the force exerted by gravity, $\vec{\tau}$ is the viscosity tensor, U is the internal energy density,
$\Phi$ is the dissipation function, $k$ is the thermal conductivity, and $\dot{q}$ is the thermal source.
The computational resources needed for a full 3D flow calculations are significant and 
are usually not necessary for an accurate calculation of the thermal solution within the 
pellets and clad.  This allows us to use a two-equation approximation in which we assume that
the coolant flow is only in the z-direction and neglect thermal diffusion between the channels.
Assuming steady-state this reduces to:
\begin{align}
& \frac{\partial \rho v}{\partial z} = 0  \\
& \frac{\partial \rho v^2}{\partial z} + \frac{\partial p}{\partial z} + g = 0 \\
& \frac{\partial U v}{\partial z} = -p \frac{\partial v}{\partial z} +
    \frac{\partial}{\partial z} \left( \nabla k(T) \frac{\partial T}{\partial z} \right)
    + \dot{q}
\end{align}
We have two different models for solving these equations that can be used interchangeably.  
The first model described in \ref{sec:single_eq_flow} uses a reduced empirical
model that solves a single axial equation using a simple finite difference scheme and
is frequently sufficient for many calculations within the nuclear engineering community.  The second model described in 
\ref{sec:subchannel_flow} is a more complex two equation model that is 
used when subchannel temperatures and densities are required.

\subsection{Single EquationFlow}
\label{sec:single_eq_flow}

A standard reduced uniaxial model based on conservation of energy in the coolant 
(Equation \ref{coolantCOE}) with a given mass flux ($G$) and a specific heat 
capacity at constant pressure ($C_p$) is employed. 
\begin{equation}
G C_{p} \frac{d T_f}{d z} + \left < h_{f} \right > T_f = \left < h_{f} T_c \right >, 
\label{coolantCOE} 
\end{equation}
where $\left < \bullet \right >$ is the integral of $\bullet$ over the heated perimeter 
of the outer surface of the clad and $T_f$ is the bulk coolant temperature.

The conservation of coolant energy is solved on a 1D domain using a finite difference scheme. 
This 1D domain is divided into $N$ equidistant grid points leading to a set of coupled equations with the $i$-th equation given by
\begin{equation}
 T_{f}^{i} - T_{f}^{i-1} + \frac{4 h_{f}(z)}{C_{p}(T_{f}^{i})G D_{e} } (T_{f}^{i}-<T_{c}^{i}> ) dz = 0
\end{equation}

where $h_f$ is the Dittus-Boelter \cite{frapcon_manual} film conductance given by
\begin{equation}
\label{Dittus-Boelter} 
h_f = (0.023k/D_{e}) Re^{0.8}Pr^{0.4}
\end{equation}
with $Re$ the Reynolds number and $Pr$ the Prandtl number.

\subsection{Subchannel Equations}
\label{sec:subchannel_flow}

In this section we describe the model that approximates the distribution of flow, 
pressure, and temperature within a channel (the space between adjacent fuel rods) 
as uniaxial in the vertical direction and utilizes empirically-derived friction 
factors to account for the additional complexities.  In our subchannel model we 
solve the 1D set of 2 equations described in section \ref{sec:Coolant}.  
Solving these equations require two independent variables per grid point
and we choose enthalpy and pressure. 
Note that the internal energy density is related to the enthalpy density through $U = h-p$.

\begin{align}
& \frac{\partial p}{\partial z} + \rho v \frac{\partial v}{\partial z} + g = 0 \\
& \frac{\partial h v}{\partial z} = v \frac{\partial p}{\partial z} +
    \frac{\partial}{\partial z} \left( k(T) \frac{\partial T}{\partial z} \right)
    + \dot{q}
\end{align}

At each axial layer, a simplified model where the crossflow terms are neglected thus eliminating 
the conservation of mass and lateral momentum equations is employed. This results in conservation 
of energy and axial momentum equations with specific enthalpy and pressure as variables using 
complex material models for the temperature, specific enthalpy, density, specific heat capacity, 
thermal conductivity, viscosity, and surface tension \cite{CobraTF_manual}.

The following finite difference form of conservation of energy and axial momentum equation are given:
\begin{multline}
	m(h_{i,j^+}-h_{i,j^-})
	-\Delta z_j\sum\limits_{r\in R(i)}(1+\gamma_{i,r})P^{heat}_r\psi_{i,r}q''_{c,j}\\
	+\Delta z_j\sum\limits_{k\in K(i)}w^t_{k,j}(h_{i,j}^*-h_{n,j}^*)
	+\Delta z_j\sum\limits_{k\in K(i)}C_{k,j} s_k(T_{i,j}-T_{n,j})=0. \\
\end{multline}
\begin{multline}
	m(u_{i,j^+}-u_{i,j^-})
	+A_i(p_{i,j^+}-p_{i,j^-})
	+\frac{g A_i\Delta z_j\cos\theta}{\nu_{i,j}^*}\\
	+\frac{1}{2A_i}\left(\frac{\Delta z_j f_{i,j}}{D_i}+\kappa_{i,j}\right)\left|m\right|m\nu_{i,j}^*
	+C^t\Delta z_j\sum\limits_{k\in K(i)}w_{k,j}^t(u_{i,j}-u_{n,j})=0.
\end{multline}

The heat flux $q''_{c,j}$ is computed using a convective heat transfer coefficient $h^{conv}_{i,j}$ and the temperature difference between the clad surface and the temperature of the flow in the center of the subchannel:
\begin{equation}
\label{flowHeatFlux} 
q''_{c,j}=h^{conv}_{i,j}(<T_{c}^{j}>-T_{i,j}).
\end{equation}
The convective heat transfer coefficient is related to the Nusselt number $Nu_{i,j}$:
\begin{equation}
Nu_{i,j}=\frac{h^{conv}_{i,j}D_i}{k_{i,j}}
\end{equation}
The Nusselt number will vary depending on the turbulence of the flow, so different models were developed for laminar and turbulent flow. To avoid convergence issues due to the discontinuity between the models, the effective heat transfer coefficient is taken as the maximum of the laminar heat transfer coefficient $h^\ell_{i,j}$ and the turbulent heat transfer coefficient $h^t_{i,j}$:
\begin{equation}
h^{conv}_{i,j}=\max(h^\ell_{i,j},h^t_{i,j}),
\end{equation}
where the laminar heat transfer coefficient is evaluated with $Nu_{i,j}=8.0$:
\begin{equation}
h^\ell_{i,j} = 8.0\frac{k_{i,j}}{D_i},
\end{equation}
and the turbulent heat transfer coefficient is calculated as
\begin{equation}
h^t_{i,j} = 0.023 Re^{0.8}_{i,j} Pr^{0.4}_{i,j}\left(\frac{k_{i,j}}{D_i}\right),
\end{equation}
which uses the well-known Dittus-Boelter correlation for Nusselt number [\ref{Dittus-Boelter}]:
\begin{equation}
Nu_{i,j} = 0.023 Re^{0.8}_{i,j} Pr^{0.4}_{i,j},
\end{equation}
where the Reynolds number and Prandtl number have their usual definitions:
\begin{equation}
Re_{i,j} = \frac{\rho_{i,j} u_{i,j} D_i}{\mu_{i,j}} , Pr_{i,j} = \frac{c^p_{i,j}\mu_{i,j}}{k_{i,j}}
\end{equation}

As boundary conditions, axial inlet mass flow rates, inlet temperature and outlet pressure are selected. 
\section{Radiation Transport Model}
\label{sec:rad_trans_model}
Nuclear fuel simulation requires a heat source and a heat sink.
The heat sink is approximated with a closed-channel coolant flow model and the 
heat source is generated within an operating nuclear reactor primarily as a result of
neutron-induced fission in fuel (primarily Uranium-235 in most reactors).
The distribution of neutrons distribution requires the solution to the Boltzmann 
transport equation \cite{LewisMiller}.  This is most often modeled using the
$k$-eigenvalue form of the Boltzmann transport equation given by
 \begin{equation}
 \begin{aligned}
  \hat{\Omega} \cdot \nabla \psi(\hat{\Omega},E) &+ \sigma(E) \psi(\hat{\Omega},E) \\
    &= \int_0^\infty \, dE' \int_{4\pi} d \hat{\Omega}'
     \sigma_s(\hat{\Omega}'\to\hat{\Omega},E'\to E) \psi(\hat{\Omega}',E') \\
     &+ \frac{1}{k} \chi(E) \int_0^\infty \, dE' \int_{4\pi} d \hat{\Omega}'
     \nu \sigma_f(E') \psi(\hat{\Omega}',E') \:, \label{eq:boltzmann}
 \end{aligned}
 \end{equation}
where $\hat{\Omega}$ is the direction of particle travel; $E$ is the particle energy;
$\psi$ is the angular flux distribution; $k$ is the multiplication factor of the system;
$\sigma$, $\sigma_s$, and $\sigma_f$ are
the total, scattering, and fission cross sections, respectively; $\chi$ is the fission
energy spectrum; and $\nu$ is the number of neutrons produced per fission.  For a given
flux distribution, the power distribution can be written as
\begin{equation}
 P = \int_0^\infty \, dE' \int_{4\pi} d \hat{\Omega}'
     \kappa \sigma_f(E') \psi(\hat{\Omega}',E') \:, \label{eq:psi_power}
\end{equation}
where $\kappa$ is the energy release per fission event.

The Denovo radiation transport code \cite{evans_2010} offers a variety of spatial
discretizations solving the discrete ordinates \cite{LewisMiller} form of Eq.~\ref{eq:boltzmann}
in parallel on Cartesian meshes.  Denovo has demonstrated excellent scalability to high
performance computing environments \cite{jarrell_2013}.  Nuclear data is generated using
the XSProc module of the SCALE package \cite{SCALE}.

Two approaches are available for transferring the power distribution from Denovo to AMP
are possible.  The first is a direct point-wise mapping of the Denovo solution onto the
Gauss points of the AMP finite element basis functions.  Interpolation to a point within
Denovo can be piecewise constant, linear, or tri-linear depending on the spatial discretization
used.  The second approach uses a polynomial expansion of the power distribution within each
cylindrical fuel rod (Zernike polynomials in the $x-y$ plane and Legendre polynomials in
the axial direction) to smooth out artifacts of the Denovo Cartesian mesh, allowing a coarser
spatial mesh to be used \cite{hamilton_2012}.  In both cases, conservation of the globally
integrated power is enforced by normalizing the distribution before and after mapping the
power onto the AMP mesh.

\section{Oxide Model:} 
\label{sec:OxideModeling}
\begin{figure}[h1]
\begin{center}
\begin{tikzpicture}
    \draw (0,0) -- (0,2);
    \draw (2,0) -- (2,2);
    \draw (4,0) -- (4,2);
    \draw (6,0) -- (6,2);
    \node (pncan) at (-1,1) {Coolant};
    \node (pncan) at (1,1) {Oxide};
    \node (pncan) at (3,1) {alpha};
    \node (pncan) at (5,1) {beta};
\end{tikzpicture}
\caption{Sample layers for oxide growth}
\label{fig:oxideRegions}
\end{center}
\end{figure}
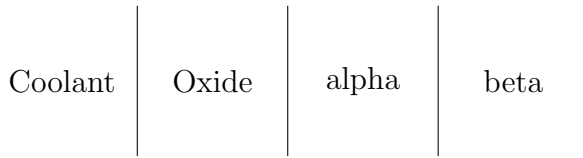

The formation of an oxide layer on the surface of a nuclear fuel rod can interfere with 
thermal conduction to the coolant and create additional stress within the clad.  
As shown in Figure \ref{fig:oxideRegions}, the material regions of interest can be divided into 4 regions, 
a coolant region which is the source of the oxygen for oxide growth, the oxide layer itself
which consists of Zirconium oxide, an oxygen rich alpha phase, and a normal beta phase region \cite{oxide_1}.

Within each layer the oxygen concentration is governed by thermal diffusion. 
Since the thickness of the oxide and alpha layers is much smaller 
than the other physical scales in the fuel rod, the oxide growth is modeled
at each point, $\mathbf{x}$, on the surface of the clad by independent 1D models:
\begin{equation}
    \displaystyle \frac{\partial C(\mathbf{x},t)}{\partial t} = 
    \frac{\partial}{\partial x}\left[D(\mathbf{x})\frac{\partial C(\mathbf{x},t)}{\partial x}\right]
\end{equation}
Solving for the oxygen concentration then reduces to solving in each region
subject to the appropriate boundary conditions.  Note that the boundaries between the different 
phases move as the different layers grow.

\noindent{\it Oxide layer growth: }The oxide layers are moving domains in which the growth of each layer is given by the 
difference between the oxygen flux between the layers.  
Associated with each interface is the associated velocity of that interface.  
For example, the velocity of the oxide-alpha layer $v_{ox,\alpha}$ is given by:
\begin{equation}
\displaystyle v_{ox,\alpha} = \frac{J_{ox}(x_{ox,\alpha})-J_{\alpha}(x_{ox,\alpha})}{C_{ox,\alpha}-C_{\alpha,ox}}
\end{equation}
where $J_{ox}(x_{ox,\alpha})$ and $J_{\alpha}(x_{ox,\alpha})$ are the flux of oxygen in the oxide and $\alpha$
layers evaluated at the oxide interface and $C_{ox,\alpha}$ and $C_{\alpha,ox}$ are the oxygen 
concentrations at the boundaries of the oxide-alpha interface in the oxide and alpha layers.

\noindent{\it Boundary conditions: }The boundary conditions between the different layers are relatively simple.  At each interface, the 
oxygen concentration can be evaluated using the equilibrium value for the given phase.  For example,
at the oxide-coolant interface, the oxygen concentration is assumed to be 1.511 $g/cm^3$, derived from 
stoichiometry.  

\noindent 
The oxygen concentration in the oxide at the oxide-alpha interface is modeled by $C = 1.517-7.5*10^{-5}*T$
\cite{oxide_1}.
The oxygen concentration in the alpha layer at the alpha-oxide interface is a fixed 29$\%$ (atomic density) 
or $\sim$0.45 37 $g/cm^3$ \cite{oxide_2}.
The oxygen concentration in the alpha layer at the alpha-beta interface is calculated using the equilibrium 
concentration.  This can be expressed as \cite{oxide_2}:
\begin{align*}
C &= -0.2263+0.0649*\sqrt{\frac{T}{63.385}-16.877} \;\;\;\;\;\;\; &T >= 1123 \;K \\
C &= 0 \;\;\;\;\;\;\;  &otherwise
\end{align*}

\begin{figure}[h2]
\begin{center}
\begin{tikzpicture}
    \draw[line width=2mm] (0,0) -- (0,2);
    \draw (2,0) -- (2,2);
    \draw (4,0) -- (4,2);
    \draw (6,0) -- (6,2);
    \draw (8,0) -- (8,2);
    \draw[line width=2mm] (10,0) -- (10,2);
    \node (pncan) at (0,-0.4) {$v_0$};
    \node (pncan) at (10,-0.4) {$v_N$};
    \node (pncan) at (4,-0.4) {$v_i$};
    \node (pncan) at (6,-0.4) {$v_{i+1}$};
    \node (pncan) at (3,1) {$C_{i-\frac{1}{2}}$};
    \node (pncan) at (5,1) {$C_{i+\frac{1}{2}}$};
    \node (pncan) at (7,1) {$C_{i+\frac{3}{2}}$};
\end{tikzpicture}
\caption{Sample layers for numerical form}
\label{fig:oxideFiniteDifference}
\end{center}
\end{figure}
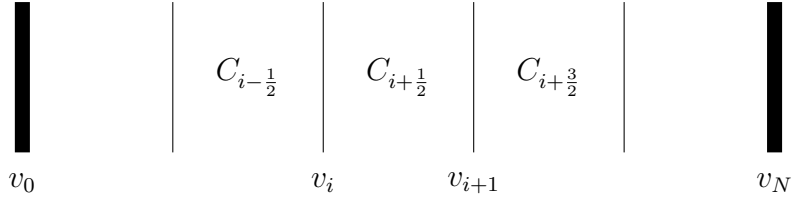

\noindent{\it Discretization: }Since we are primarily conserned with the size of the oxide layers, we choose
to solve the differential equation using finite difference in a moving frame.
\begin{equation}
  \frac{d C(\mathbf{x},t)}{d t} = \frac{\partial}{\partial x}\left[D(\mathbf{x})\frac{\partial C(\mathbf{x},t)}{\partial x}\right] + v\cdot\frac{\partial C(\mathbf{x},t)}{\partial x}
\end{equation}
We divide the space into N uniformly spaced regions of size $h = (x_N-x_0)/N$.
Since each zone is moving at a velocity $v_i$, we can apply the convective derivative
to get the diffusion equation in this moving frame.  
We can follow the boundaries of the oxide layer by choosing the velocity 
at the boundaries to match the oxide growth rate.  
For example consider the layers shown in Figure \ref{fig:oxideFiniteDifference}.  
It's left boundary is moving at a velocity of $v_1$ and it's right boundary is moving at a velocity of $v_2$.

\noindent

We want to use a conservative scheme with upwinding for the convective term.  Assuming $v_{1+1/2}\geq 0$:
\begin{align*}
  & \frac{d C_{i+\frac{1}{2}}}{d t} = \frac{1}{h^2}\left[ D_{i+1}\left(C_{i+\frac{3}{2}}-C_{i+\frac{1}{2}}\right) - D_{i}\left(C_{i+\frac{1}{2}}-C_{i-\frac{1}{2}}\right) \right] 
     + \frac{v_{i+\frac{1}{2}}}{h}\left(C_{i+\frac{3}{2}}-C_{i+\frac{1}{2}}\right) \\
  & v_{i+\frac{1}{2}} = v_0+\frac{i+\frac{1}{2}}{N}\left(v_N-v_0\right)
\end{align*}

Rewriting:
\begin{align*}
  & \frac{d C}{d t} = F(C,v,h) \\
  & F(C,v,h) = \frac{1}{h^2}\left[D_{i+1}\left(C_{i+\frac{3}{2}}-C_{i+\frac{1}{2}}\right)-D_{i}\left(C_{i+\frac{1}{2}}-C_{i-\frac{1}{2}}\right)\right] 
    + \frac{v_{i+\frac{1}{2}}}{h}\left(C_{i+\frac{3}{2}}-C_{i+\frac{1}{2}}\right) \\
\end{align*}
We can then apply the Crank-Nicholson method:
\begin{flalign*}
  \frac{C^{n+1}-C^{n}}{\Delta t} &= \frac{1}{2}\left[F\left(C^{n+1},v^{n+1},h^{n+1}\right) + F\left(C^n,v^n,h^n\right)\right] \\
  h^{n+1} &= \left(x_N^{n+1}-x_0^{n+1}\right)/N \\
  x_0^{n+1} &= x_0^n + v_0^n\Delta t + 0.5\left(v_0^{n+1}-v_0^n\right)\Delta t^2 \\
  x_N^{n+1} &= x_N^n + v_N^n\Delta t + 0.5\left(v_N^{n+1}-v_N^n\right)\Delta t^2 \\
\end{flalign*}
If we assume $v^{n+1}$ is known (it is actually calculated from $C^{n+1}$), 
then the system becomes a standard linear system of equations.  
The resulting matrix is banded and can be solved through direct solves using LAPACK.

\noindent {\it Time-Step Control: }Careful control of the time step is necessary to produce an accurate answer.  
A first limitation of the time step is the calculation of $v^{n+1}$.  
While it is possible to create a non-linear system that solves for $v^{n+1}$, 
the other time step requirements will make this work unnecessary.  
Instead we will assume $v^{n+1} \approx v^{n}$.  With this assumption we 
need a time step that will ensure the change in $v$ is small.  
The second restriction comes from the convective term.  
To ensure the proper error, we need to limit the matrix norm (for this term) to $\sim 1$.  
This gives us the condition $\Delta t\leq\frac{h}{v_i}$.

\end{document}